\documentclass[12pt,letterpaper]{article}
\usepackage{jheppub} 
\usepackage{amsmath,amsthm,amsfonts,amssymb,amscd}
\usepackage{tikz}
\usepackage{subcaption}

\makeatletter
\newsavebox{\@brx}
\newcommand{\llangle}[1][]{\savebox{\@brx}{\(\m@th{#1\langle}\)}%
	\mathopen{\copy\@brx\mkern2mu\kern-0.9\wd\@brx\usebox{\@brx}}}
\newcommand{\rrangle}[1][]{\savebox{\@brx}{\(\m@th{#1\rangle}\)}%
	\mathclose{\copy\@brx\mkern2mu\kern-0.9\wd\@brx\usebox{\@brx}}}
\makeatother

\definecolor{darkred}{rgb}{0.4,0,0}
\definecolor{lazuli}{rgb}{0.0, 0.06, 0.54}

\newcommand{\vev}[1]{\left\langle #1 \right\rangle}
\newcommand{\nn}{\nonumber}

\def\ket#1{|#1\rangle}

\newcommand{\mcO}{\mathcal{O}}
\newcommand{\td}{\text{d}}

\newcommand{\zb}{\Bar{z}}
\newcommand{\hb}{\Bar{h}}
\newcommand{\oket}[1]{  \left. \vert #1 \right) }

\title{Krylov Complexity in Periodically Driven CFTs and Critical Fermions}
\date{\today}

\author[a]{Ankit Gill,}
\author[b]{Anurag Sarkar}

\affiliation[a]{ Department of Physics, Indian Institute of Technology, Kanpur, UP 208016, India.}
\affiliation[b]{ Mandelstam Institute for Theoretical Physics, School of Physics, University of the Witwatersrand, Johannesburg, WITS 2050, South Africa.}

\emailAdd{ankitgill20@iitk.ac.in}
\emailAdd{anurag.sarkar@wits.ac.za}

\abstract{We study Krylov construction in periodically driven conformal field theories and their lattice realisations via critical fermions. Two types of driving are considered: a square-wave drive and a continuous sinusoidal drive. Using the Arnoldi construction, we examine Arnoldi coefficients and return amplitudes in periodically driven conformal field theories in the heating and non-heating phases. In the heating phase, the Arnoldi coefficients approach unity exponentially; in contrast, in the non-heating phase, they exhibit oscillatory behaviour. For the lattice realisations, we further analyse the Krylov complexity of the correlation matrix, quasi energy level statistics, and the graph structure induced by the Floquet operator. Although the two drives exhibit similar Krylov growth on the CFT side, their lattice realisations exhibit markedly different spectral and graph signatures, indicating distinct mechanisms governing the transition between the heating and non-heating phases.}

\begin{document}
	
\maketitle
	

\section{Introduction}
Quantum dynamics in many-body systems refers to the evolution of quantum degrees of freedom under a Hamiltonian~\cite{Polkovnikov:2010yn,Nandkishore:2014kca}. As the system evolves unitarily under a generic local Hamiltonian, an initially simple product state can become increasingly complex, developing extensive multipartite entanglement~\cite{Calabrese:2005in,Calabrese:2007rg,2019PhRvR...1c3067L, Kim:2013etb,AnthonyChen:2023bbe}. In this process, local information spreads across the system, leading to information scrambling~\cite{Maldacena:2015waa,Shenker:2013pqa,Xu:2022vko,Hosur:2015ylk,Nahum:2017yvy}, a phenomenon in which initially localized quantum information becomes delocalised into highly nonlocal many-body correlations. Such evolution can lead to thermalisation, consistent with the Eigenstate Thermalisation Hypothesis~\cite{PhysRevA.43.2046,Srednicki:1994mfb,2008Natur.452..854R,Popescu:2006rhr,2008PhRvL.101s0403R}, which posits that expectation values of local operators in highly excited eigen-states effectively take thermal values. However, integrable systems~\cite{Rigol:2016itf,Essler:2016ufo} and constrained models~\cite{Turner:2017fxc,Serbyn:2020wys} can avoid this behaviour, thereby exhibiting non-ergodic and non-thermal dynamics.

Building on this, periodically driven systems extend the framework by introducing explicit time dependence into the Hamiltonian. Such driving can serve as a powerful technique for engineering nontrivial effects and phases absent in equilibrium systems. Floquet theory provides a framework for describing periodically driven systems in terms of stroboscopic evolution~\cite{PhysRev.138.B979}. It maps the dynamics onto an effective time-independent Hamiltonian that governs the system at discrete time intervals~\cite{PhysRevA.7.2203,2015AdPhy..64..139B}. 

These periodically driven systems, as a result, exhibit a rich range of genuinely nonequilibrium phenomena, including dynamical localisation~\cite{PhysRevA.77.010101,PhysRevB.96.144301,Tamang:2020dqy,Aditya_2023}, Floquet-engineered topological phases~\cite{2011NatPh...7..490L, 2010PhRvB..82w5114K, PhysRevLett.114.056801,Rudner:2019adp}, and discrete-time crystals~\cite{Zhang:2016kpq}. Such effects have been realised experimentally across diverse platforms, including ultracold atoms~\cite{Bordia:2016owe,Wintersperger:2020rqb}, trapped ions~\cite{Zhang:2016kpq}, and superconducting qubits~\cite{Frey:2021iix}, highlighting the versatility of driven quantum systems for exploring and engineering nonequilibrium many-body physics.

In this work, we focus on (1+1)-dimensional conformal field theories and critical fermions on one-dimensional lattices. Periodically driven CFTs provide an analytically tractable framework for exploring nonequilibrium criticality. Recent works~\cite{Wen:2018agb,Han:2020kwp} showed that periodic driving in CFT can be described via Mobius transformations, leading to distinct heating and non-heating phases. In Floquet CFT, the non-heating phase is the regime in which the return amplitude oscillates, and the energy absorption remains bounded. In contrast, in the heating phase, the return amplitude decays exponentially, and the system continuously absorbs energy under periodic driving, eventually approaching an infinite-temperature-like state \cite{Wen:2018agb}. This framework has also been extended to quasiperiodic and random driving in (1+1)d Floquet CFTs~\cite{Wen:2020wee}, where the dynamics exhibits fractal phase structures. In~\cite{Lapierre:2020ftq,Fan:2020orx}, the formalism is extended beyond $\mathrm{SL}(2)$ (Mobius) deformations to arbitrary smooth spatial deformations of the Hamiltonian density, where the resulting phase structure---encompassing heating, non-heating, and their subphases---is determined by the fixed points of operator evolution. Other works have focused on periodically driven perturbed CFTs, such as the sine-Gordon model~\cite{Bajnok:2021twm}, continuous drive protocols~\cite{Das:2025wjo}, and recent extensions to exactly solvable and higher-dimensional Floquet CFTs~\cite{Das:2023xaw,Das:2021gts} that exhibit rich dynamical phase structures.

To better understand the dynamics in these models, a range of diagnostics has been employed in the literature for Floquet CFTs and periodically driven critical fermions, including the Floquet quasienergy spectrum~\cite{Bajnok:2021twm}, as well as entanglement entropy, return probability, and energy density~\cite{Wen:2018agb,Das:2021gts,Fan:2020orx,Berdanier:2017kmd}. In addition, equal- and unequal-time two-point correlation functions~\cite{Andersen:2020xvu} and out-of-time-order correlators~\cite{Das:2022jrr,Khetrapal:2022dzy,Lapierre:2024lga} and Modular Hamiltonians~\cite{Das:2024vqe} have been considered.

Among these diagnostics, Krylov complexity~\cite{Parker:2018yvk} has proven to be a reliable framework for characterising operator growth and quantum chaos. The notion of spread complexity~\cite{PhysRevD.106.046007} provides a complementary state-based approach, which will be the focus of this work. The Krylov-based approaches offer an intuitive way to describe the dynamics of quantum states/operators as they spread into a hierarchy of increasingly complex states/operators. The recursion method builds this hierarchy as an orthogonal Krylov basis, where each step corresponds to a state of increasing complexity. In this picture, the K-complexity is measured by the average position of the quantum state/operator on that basis. Recent studies have further developed Krylov methods for probing chaotic dynamics, operator growth, and transitions between integrable and chaotic regimes, including random-matrix theory for K-complexity growth and saddle-dominated scrambling \cite{Kar:2021nbm,Bhattacharjee:2022vlt,Huh:2023jxt,Baggioli:2024wbz,Camargo:2024deu}. For a more comprehensive review on methods in Krylov space, see~\cite{Nandy:2024evd,Rabinovici:2025otw}.

Specifically in the context of conformal field theory, operator growth in 2d conformal field theories and holographic models has been studied~\cite{Dymarsky:2021bjq,Kundu:2023hbk}. More recently, Krylov complexity has been studied in deformed CFTs, ~\cite{Chattopadhyay:2024pdj}, as well as in $\mathrm{SL}(2,\mathbb{R})$-deformed Floquet-like Hamiltonians, where it was shown to capture heating and non-heating dynamical phases~\cite{Malvimat:2024vhr}. This framework has also been applied to quantum field theories in~\cite{Adhikari:2022whf,Avdoshkin:2022xuw,Camargo:2022rnt,He:2024xjp}.

Further developments have extended the Krylov-based approach to driven quantum systems~\cite{Nizami:2023dkf,Nizami:2024ltk,PhysRevLett.134.030401,Grabarits:2026hjz}. Subsequent works explored the role of Krylov complexity in characterising Floquet thermalisation and prethermal behaviour~\cite{2024arXiv240418052Q}, as well as Trotterized Floquet evolutions in~\cite{PhysRevB.111.014309}. More recently, these ideas have been further developed using the theory of orthogonal polynomials to classify ergodic Floquet systems~\cite{Kolganov:2024nzi} and for ergodic and dynamically frozen phases using Krylov-space diagnostics~\cite{2025arXiv251019824S}. The dependence of Krylov complexity growth and saturation on the choice of initial operator and state has also been investigated in driven settings~\cite{PG:2025ixk}.

In this work, we will examine the K-complexity in CFT within the $\mathrm{SL}(2,\mathbb{R})$ sector and in critical free fermions under two Floquet drive protocols. The first corresponds to a square-wave drive, where the system is periodically driven by a deformed sine-squared Hamiltonian in discrete steps~\cite{Wen:2018agb}. The second drive is the continuous sinusoidal drive~\cite{Das:2021gts}. We construct the Krylov Basis by orthogonalising the basis constructed by the repeated application of the $U_F = e^{-i H_{F} T}$ ~\cite{PhysRevB.104.195121,Nizami:2023dkf,Nizami:2024ltk,PhysRevB.111.014309}, where the $H_F$ is the effective Floquet hamiltonian. In such a basis, $U_F$ assumes an upper Hessenberg form (i.e.\ all matrix elements are zero below the first sub-diagonal), the non-zero elements $h_{i,j}$ are identified as the Arnoldi coefficients \cite{Nizami:2023dkf,PhysRevB.104.195121}. These coefficients govern the transitions between states of different complexity levels. The sub-diagonal elements $h_{n,n-1}$ are particularly important, as they control the coupling between successive Krylov basis states generated by repeated applications of $U_F^{n}$. For both protocols, we find that in the heating phase, $h_{n,n-1} \rightarrow 1$ as n increases, and in the non-heating phase, they show oscillatory behaviour.

We also perform lattice simulations of the CFT dynamics and find agreement in both return amplitude and Arnoldi coefficients for smaller values of $n$. In the square-wave drive, the crossover between the heating and non-heating regimes is captured by the average level-spacing ratio $\langle r \rangle$ of the quasienergies of $U_F$. We find that $\langle r \rangle$ fluctuates around the Wigner--Dyson value of $0.53$ in the heating phase, while in the non-heating phase it fluctuates around the Poisson value of $0.386$. In contrast, for the continuous drive, $\langle r \rangle$ shows erratic fluctuations in the heating phase, while in the non-heating phase it varies smoothly but does not saturate to either the Wigner-Dyson or Poisson values. To further analyze the unitary matrix $U_F$, we construct the stochastic transition matrix $W_{ij} = |(U_F)_{ij}|^2$ \cite{2001JPhA...34.8485T,2001JPhA...34L.319B,Pakonski:2001uok,Gualtieri:2020lug,Bressanini:2022qrr,Severini:2005gbh}, which defines a weighted graph over lattice sites. This allows us to reinterpret the dynamics as a classical Markov process and probe its connectivity through graph-theoretic measures. In particular, the Fiedler value~\cite{Fiedler1973AlgebraicCO,chung1997spectral} provides a quantitative diagnostic of how the effective connectivity evolves across different driving regimes.

\section{Review of K-Complexity} \label{sec:K-complexity}

We start our discussion with a brief review of the Krylov basis construction.
Consider the autocorrelation function.

\begin{equation}
\big( O_{0} | O(t) \big) = Tr(O_{0}^{\dagger} e^{-i  H t } O_{0} e^{i  H t }  ),
\end{equation}

\noindent by Taylor series expansion and $\mathcal{L} = [H,.]$:

\begin{equation}
O(t) = \sum_{n =0 }^{\infty} \frac{1}{n!} \mathcal{L}^{n}( O_{0})(i t)^{n}.
\end{equation}
 
The Krylov basis is constructed from by orthonormalisation of the basis \cite{Parker:2018yvk} $$\{ O_{0}, \mathcal{L} O_{0} ,\mathcal{L}^2 O_{0},\dots\}$$ by Lanczos Algorithm by taking $| K_{0} ) = | O_{0} )$ as first Krylov vector. Then the consecutive Krylov vectors are generated recursively as,

\begin{equation}
 | K_{n+1} ) = \frac{1}{b_{n+1}} \mathcal{L} |K_{n} )  - b_{n-1} | K_{n-1} ) ,
\end{equation}

\noindent where $b_{n}$ are fixed such that $(K_{n} | K_{n})  = 1$ are normalised. K- complexity can be defined as the 

\begin{equation}
 C_{K} = \sum_{n} n | (K_{n} | O_{t} ) |^2.
\end{equation}

For a periodically driven system, the next Krylov vector is obtained by applying the Unitary Floquet Operator $U_{F}$ \cite{PhysRevD.106.046007,Nizami:2023dkf}, followed by orthonormalisation against all previous Krylov vectors (known as the Arnoldi Iteration). The Krylov basis is thus constructed as $\big\{ O_{0}, U_{F}^{\dagger} O_{0} U_{F}, \dots \big\}$, where $U_{F} |K_{n}) = U_{F}^{\dagger} K_{n} U_{F}$.

\begin{align}
 |K_{n}) = \frac{1}{h_{n,n-1}} [ U_{F} |K_{n-1}) - \sum_{j=0}^{n-1} h_{j,n-1} |K_{j})   ],
    \label{kconst}
\end{align}

The inner product $h_{j,k} = (K_j | U_F | K_k)$ is defined by a suitable choice (e.g., $(A|B) \equiv Tr(A^{\dagger}B)$) and $h_{n,n-1}$ is fixed so that $(K_n|K_n)=1$, ensuring normalization. Operator complexity is then expressed in this Krylov basis as:
\begin{align}
 C_{K}(j) = \sum_{n=0} n | (K_{n} | U_{F}^{j} | K_{0})|^2 .
\end{align}

In this Krylov basis, $U_{F}$ becomes an upper triangular matrix with elements $h_{k,j}$, where unitarity ensures $|h_{j,k}| \leq 1$. The sub-diagonal elements $h_{n,n-1}$ represent the probability amplitude for the transition from the $K_{n-1}$ operator to the $K_{n}$ operator. These elements can also serve as the probe of quantum chaos \cite{Nizami:2023dkf,Nizami:2024ltk,PhysRevB.111.014309}.

There also exsists an correspondence between the moments  $(K_{0}|U_{F}^{n} | K_{0} )$ and the $h_{j,k}$ \cite{PhysRevD.106.046007}. These matrix elements can be obtained by using the unitarity conditions of $U_{F}$ and  $(K_{0}|U_{F}^{n} | K_{0} )$. This can be particularly important in cases where the action of $U_{F} | K_{n}) $ cannot be determined or is too complicated. These coefficients are determined column-wise. In the first column, there are two coefficients $h_{0,0}$ and $ h_{1,0}$. Without loss of generality, one can take $h_{n,n-1}$ (sub-diagonal) to be real and positive. The normalisation condition of first column of $U_{F}$,

\begin{equation}
    |h_{00}|^2 + |h_{10}|^2 = 1,
\end{equation}

 \noindent determines the $h_{1,0}$. For the second column we have to determine five variables $\Re({h_{0,1}})$, $\Re({h_{1,1}})$, $ h_{2,1}$, $\Im({h_{0,1}})$, $\Im({h_{1,1}})$. These are determined by:

\begin{align}
     |h_{00}|^2 + |h_{10}|^2 &= 1, \nonumber\ \\ 
      h_{00} h_{01}^{*}  + h_{10} h_{11}^{*} &=  0,  \nonumber\\\ 
       h_{00}^{2} + h_{01} h_{10} &= (K_{0}|U_{F}^{2} | K_{0} ). 
\end{align}

\begin{figure}[h!]
    \centering
    \includegraphics[width=7cm,height=6cm]{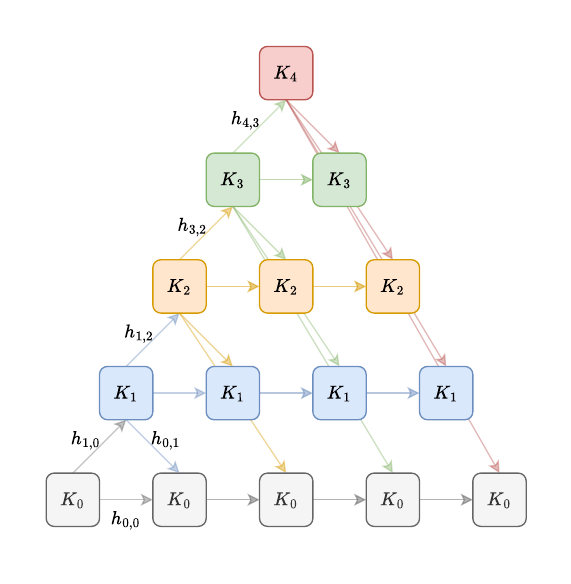}
    \caption{Schematics of Possible transitions in Krylov basis during periodic drive~\cite{Nizami:2023dkf}.}
    \label{fig:floquet ham}
\end{figure}

\noindent The complex conjugates of the second and third equations are also considered. The second equation imposes orthogonality between the first and second columns. The third equation sums all amplitudes that contribute to the second moment. Similarly, for $2n + 1$ coefficients in the $n^{th}$ column: $2(n-1)$ equations enforce orthogonality, one equation enforces normalisation, and two equations account for total transition amplitudes in the $n^{th}$ moment. Figure~\ref{fig:floquet ham} schematically presents the transitions between the Krylov vectors.

\paragraph{Relation to Operator Growth}

We begin by examining the behaviour of the sub-diagonal matrix elements of the Floquet unitary \(U_F\) expressed in the Krylov basis, namely the coefficients \(h_{n,n-1}\). These coefficients encode the spreading of the time-evolving state in Krylov space and therefore provide a useful probe of operator growth and ergodicity\cite{Nizami:2023dkf,PhysRevB.111.014309}.

In the heating phase, as the number of Floquet cycles ($n$) increases, the coefficients $h_{n,n-1}$ approach unity. Meanwhile, the other matrix elements of $U_F$ are strongly suppressed, as expected from unitarity. Thus, for large $n$, the evolution is likely to shift the state to the next Krylov basis vector, indicating rapid spreading in Krylov space. Numerically, $h_{n,n-1}$ approaches unity exponentially with $n$. This behaviour somewhat resembles maximally ergodic dynamics in random dual-unitary circuits, where the Floquet operator's matrix elements in the Krylov basis becomes dominated by its sub-diagonal structure while diagonal and upper-triangular elements vanish~\cite{PhysRevB.111.014309}.

The matrix elements \(h_{m,n}\) behave differently in the non-heating phase. Here, the diagonal and upper-triangular elements of \(U_F\) remain nonzero and oscillate as the Floquet cycle increases. Consequently, the sub-diagonal coefficients \(h_{n,n-1}\) also oscillate and does not converge to unity. This behavior results from the unitarity condition \begin{equation}\sum_{m} |h_{m,n}|^2 = 1 ,\end{equation} which must hold for every column \(n\) of the unitary matrix \(U_F\). Persistent nonzero diagonal and upper-triangular elements prevent the sub-diagonal coefficients from converging to unity, leading to oscillatory Krylov dynamics in the non-heating regime. Thus, at the \(n\)-th step, the time-evolving state has a large overlap with the previous Krylov vector and mostly transitions within the already explored Krylov subspace, rather than into new directions in the Krylov basis.


\section{Sine-square deformed (SSD) Hamiltonian}

In this section, we briefly overview the formulation of the sine-square deformed CFT Hamiltonian. We start with the undeformed Hamiltonian:
\begin{align}
    H_0 = \int_0^{L} \td x\, h(x).
\end{align}
The sine-square deformed Hamiltonian is then defined as \cite{Katsura:2011ss}:
\begin{align}
    H_{\text{SSD}}
    &= \int_0^{L} \td x\, 2\sin^2\!\left(\frac{\pi x}{L}\right) h(x) \nn = H_0 - \int_0^{L} \td x\, h(x)\cos\!\left(\frac{2\pi x}{L}\right).
\end{align}

\noindent In a two-dimensional CFT on the cylinder, the Hamiltonian reads
\begin{align}
    H_0 = \int_0^{L} \frac{\td x}{2\pi}\, T_{00}(x)
    = \int_0^{L} \frac{\td x}{2\pi}\,\bigl[T(w)+\bar T(\bar w)\bigr],
\end{align}
where $w=\tau+ix$, $\tau$ is the imaginary time and $0\le x \le L$. Under the conformal map
\begin{align}
    z=e^{2\pi w/L}, \qquad w=\frac{L}{2\pi}\log z,
\end{align}
the stress tensor transforms as
\begin{align}
    T(w)=\left(\frac{\td w}{\td z}\right)^{-2}\!\left[T(z)-\frac{c}{12}\{w;z\}\right]
    =\left(\frac{2\pi}{L}\right)^2\!\left[z^2 T(z)-\frac{c}{24}\right].
\end{align}

\begin{figure}[h!]
    \centering
    \includegraphics[width=\linewidth]{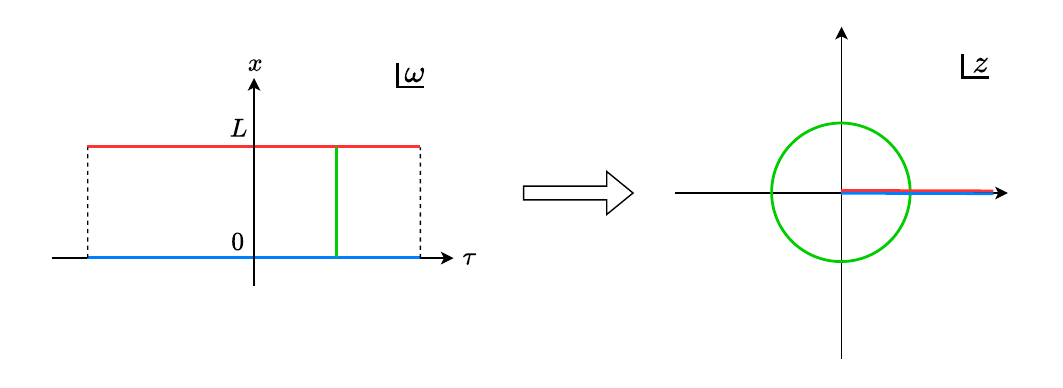}
    \caption{Conformal map from the strip to the plane.}
    \label{fig:w_to_z}
\end{figure}

\noindent Using the mode expansion $T(z)=\sum_n z^{-n-2}L_n$, one finds
\begin{align}
    H_0&=\frac{2\pi}{L}\left(L_0+\bar L_0\right)-\frac{\pi c}{6L}, \\
     H_{\text{SSD}}&=H_0-\frac12\left(H_+ + H_-\right),
\end{align}
where,
\begin{align}
    H_+ + H_- = L_1+L_{-1}+\bar L_1+\bar L_{-1}.
\end{align}

In the following sections, we will study the behaviour of the Krylov complexity in systems where the driving Hamiltonian periodically changes between $H_0$ and $H_{SSD}$, under square wave driving and continuous driving protocols.

\section{Square wave drive in CFT}

We drive the system periodically with $H_0$ and $H_{\text{SSD}}$ according to
\begin{align}
    H=
    \begin{cases}
        H_0, & (n-1)T \le t \le (n-1)T+T_0,\\
        H_{\text{SSD}}, & (n-1)T+T_0 \le t \le nT,
    \end{cases}
\end{align}
with $n\in\mathbb Z_+$ and $T=T_0+T_1$. The switch between the two Hamiltonians is taken to be instantaneous.

\begin{figure}[h!]
    \centering
    \includegraphics[width=0.8\linewidth]{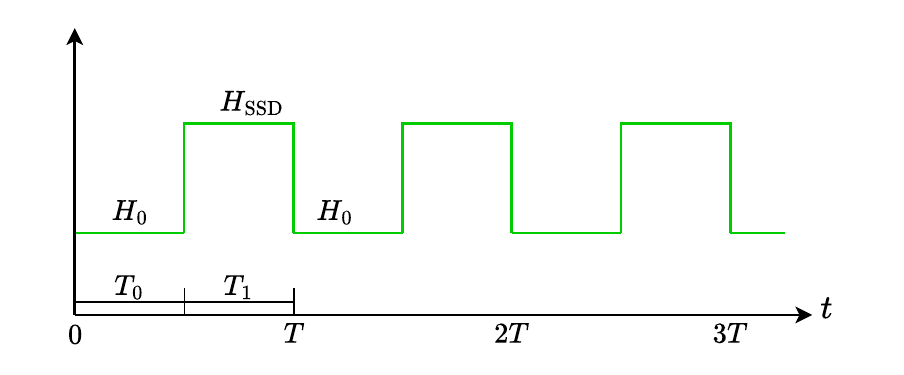}
    \caption{Schematic of the square-wave Floquet drive.}
    \label{fig:floquet_ham}
\end{figure}

We consider a $(1+1)$-dimensional CFT on the interval $x\in[0,L]$ and map the strip to the complex plane via $z=e^{2\pi w/L}$, as discussed in the previous section.

To compute operator complexity, we start from the initial operator state
\begin{align}
    \oket{\mathcal O}
    \equiv \left(\frac{2\pi}{L}\right)^{-(h+\bar h)}
    \phi(\tau\to-\infty,x=0)
    \equiv O(z=0,\bar z=0),
\end{align}
where $O$ is a primary operator of weights $(h,\bar h)$; the prefactor ensures proper normalization.

After $n$ driving cycles, the operator evolves as \cite{Wen:2018agb}:
\begin{align}
    (U_F^\dagger)^n\, O(z,\bar z)\, U_F^n
    =
    \left(\frac{\partial z_n}{\partial z}\right)^h
    \left(\frac{\partial \bar z_n}{\partial \bar z}\right)^{\bar h}
    O(z_n,\bar z_n),
\end{align}
where
\begin{align}
    U_F=e^{-iH_{\text{SSD}}T_1}e^{-iH_0T_0}.
\end{align}
The one-cycle map is a M\"obius transformation,
\begin{align}
    z_1(z)=\frac{az+b}{cz+d},
\end{align}
with
\begin{equation}
\begin{aligned}
    a &= \left(1+\frac{i\pi T_1}{L}\right)e^{i\pi T_0/L},
    \qquad
    b = -\frac{i\pi T_1}{L}e^{-i\pi T_0/L},\\
    c &= \frac{i\pi T_1}{L}e^{i\pi T_0/L},
    \qquad
    d = \left(1-\frac{i\pi T_1}{L}\right)e^{-i\pi T_0/L}.
\end{aligned}
\end{equation}
Iterating this map gives
\begin{align}
    \frac{z_n-\gamma_1}{z_n-\gamma_2}
    =
    \eta^n\frac{z-\gamma_1}{z-\gamma_2},
\end{align}
where
\begin{equation}
\begin{aligned}
    \gamma_{1,2}
    &= \frac{a-d\mp\sqrt{(a-d)^2+4bc}}{2c}, \qquad
    \eta
    = \frac{(a+d)+\sqrt{(a-d)^2+4bc}}
    {(a+d)-\sqrt{(a-d)^2+4bc}}.
\end{aligned}
\end{equation}

\subsection{Krylov-Construction}
We now set up the computation of Krylov complexity for the square wave drive, where the initial operator is taken to be,

\begin{align}
    \oket{\mcO} = \left( \frac{2\pi}{L} \right)^{ - ( h + \hb )} O(0,0).
\end{align}
Here $O(0,0)$ is a primary operator located at the origin. The evolution operator is $U_{F} \equiv e^{-i H_{SSD} T_{1} } e^{-i H_{0} T_{0} }$. To follow the construction explained in \S \ref{sec:K-complexity}, only the information about $(K_{0}|U_{F}^{n} | K_{0} )$ is required.

The quantity $(K_{0}|U_{F}^{n} | K_{0} )$ can be calculated by using the mapping $z \to z_n$ discussed earlier. One then obtains:
\begin{equation}
    \begin{aligned}
        (K_{0}|U_{F}^{n} | K_{0} ) &= \vev{ O^{\dagger} (0,0)  (U_F^\dagger)^n O(0, 0) U_F^n } \\
        &= \lim_{z, \zb \to 0} \vev{ O^{\dagger} (z,\zb)  O(z_n, \zb_n) } \left( \frac{\partial z_n}{\partial z} \right)^h \left( \frac{\partial \zb_n}{\partial \zb} \right)^{\hb} \\
        &= \left( \frac{\left(\gamma _1-\gamma _2\right){}^2 \eta ^n}{\left(\gamma _2-\gamma _1 \eta ^n\right){}^2} \right)^h \left( \frac{\left(\gamma^{*} _1-\gamma^{*} _2\right){}^2 \eta^{*} {}^n}{\left(\gamma^{*} _2-\gamma^{*} _1 \eta^{*} {}^n\right){}^2} \right)^{\hb} \lim_{z, \zb \to 0} \vev{ O (1/\zb,1/z)  O(z_n, \zb_n) } z^{-2 \hb} \zb^{-2 h} \\
        &= \left( \frac{\left(\gamma _1-\gamma _2\right){}^2 \eta ^n}{\left(\gamma _2-\gamma _1 \eta ^n\right){}^2} \right)^h \left( \frac{\left(\gamma^{*} _1-\gamma^{*} _2\right){}^2 \eta^{*} {}^n}{\left(\gamma^{*} _2-\gamma^{*} _1 \eta^{*} {}^n\right){}^2} \right)^{\hb} \lim_{\xi, \Bar{\xi} \to \infty} \vev{ O (\xi, \Bar{\xi})  O(0, 0) } (\Bar{\xi} + \zb_n)^{2 \hb} (\xi + z_n)^{2 h} \\
        &= \left( \frac{\left(\gamma _1-\gamma _2\right){}^2 \eta ^n}{\left(\gamma _2-\gamma _1 \eta ^n\right){}^2} \right)^h \left( \frac{\left(\gamma^{*} _1-\gamma^{*} _2\right){}^2 \eta^{*} {}^n}{\left(\gamma^{*} _2-\gamma^{*} _1 \eta^{*} {}^n\right){}^2} \right)^{\hb} \lim_{\xi, \Bar{\xi} \to \infty} \Bar{\xi}^{-2 \hb} \xi^{-2 h} (\Bar{\xi} + \zb_n)^{2 \hb} (\xi + z_n)^{2 h} \\
        &= \left( \frac{\left(\gamma _1-\gamma _2\right){}^2 \eta ^n}{\left(\gamma _2-\gamma _1 \eta ^n\right){}^2} \right)^h \left( \frac{\left(\gamma^{*} _1-\gamma^{*} _2\right){}^2 \eta^{*} {}^n}{\left(\gamma^{*} _2-\gamma^{*} _1 \eta^{*} {}^n\right){}^2} \right)^{\hb}.
    \end{aligned}
\end{equation}

For $h = \hb$, the above expression simplifies, and we define:
\begin{align} \label{eq:Gn}
    G_n := (K_{0}|U_{F}^{n} | K_{0} ) &= \left( \frac{\left| \gamma _1-\gamma _2 \right|{}^2 |\eta| ^n}{\left| \gamma _2-\gamma _1 \eta ^n\right|^2} \right)^{2h} .
\end{align}



\subsection{Analysis of the autocorrelation $G_n$ and the Arnoldi coefficients}

We first note that the heating and non-heating phases in the floquet dynamics can be distinguished by the value of $\Delta$, which is defined as:
\begin{align}
    \Delta = \left(\frac{\pi  T_1 }{L} \cos \left(\frac{\pi  \text{T0}}{L}\right) +\sin \left(\frac{\pi  T_0}{L}\right)\right)^2-\left(\frac{\pi  T_1}{L}\right)^2.
\end{align}
The phases are then identified as follows \cite{Wen:2018agb},
\begin{equation}
    \begin{cases}
    0 < \Delta \leq 1: ~~ &\text{Non-heating (elliptic)}, \\
    \Delta = 0: ~~ &\text{Critical (parabolic)}, \\
    \Delta < 0: ~~ &\text{Heating (hyperbolic)}.
\end{cases}
\end{equation}

\subsubsection{The heating phase}

\begin{figure}[htbp]
  \centering
  \begin{subfigure}{0.45\textwidth}
    \centering
    \includegraphics[width=\linewidth]{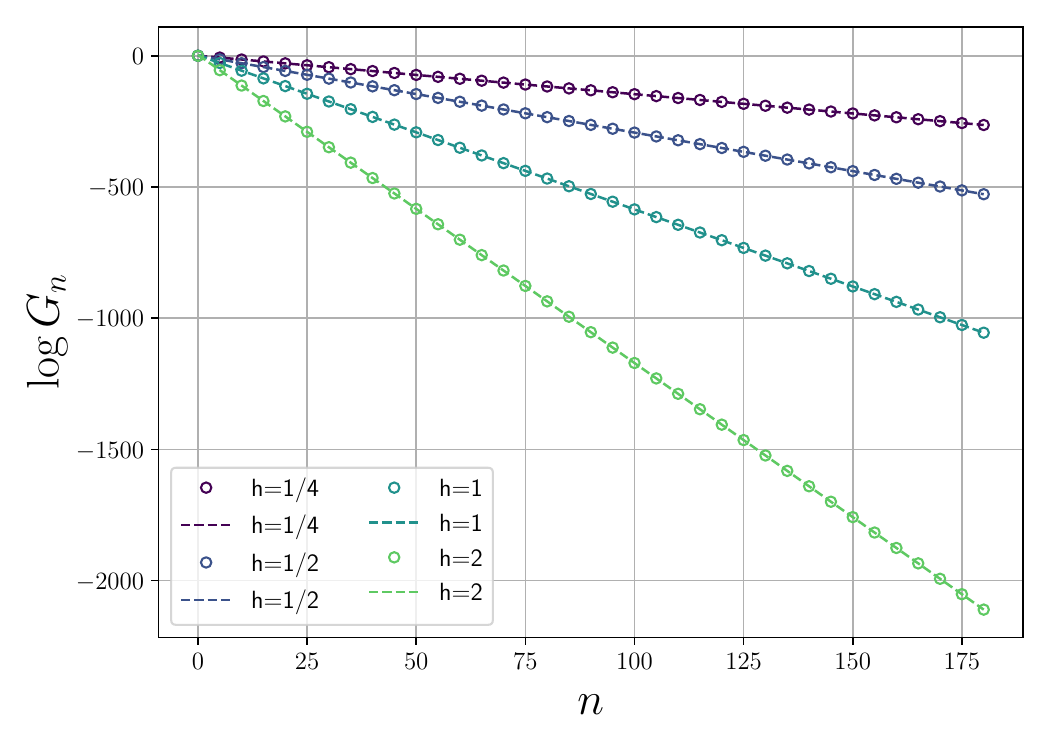}
    \caption{$(T_0/L, T_1/L) = (0.8, 0.8)$, $\eta < 1$}
    \label{fig:heating Gn 1}
  \end{subfigure}\hfill
  \begin{subfigure}{0.45\textwidth}
    \centering
    \includegraphics[width=\linewidth]{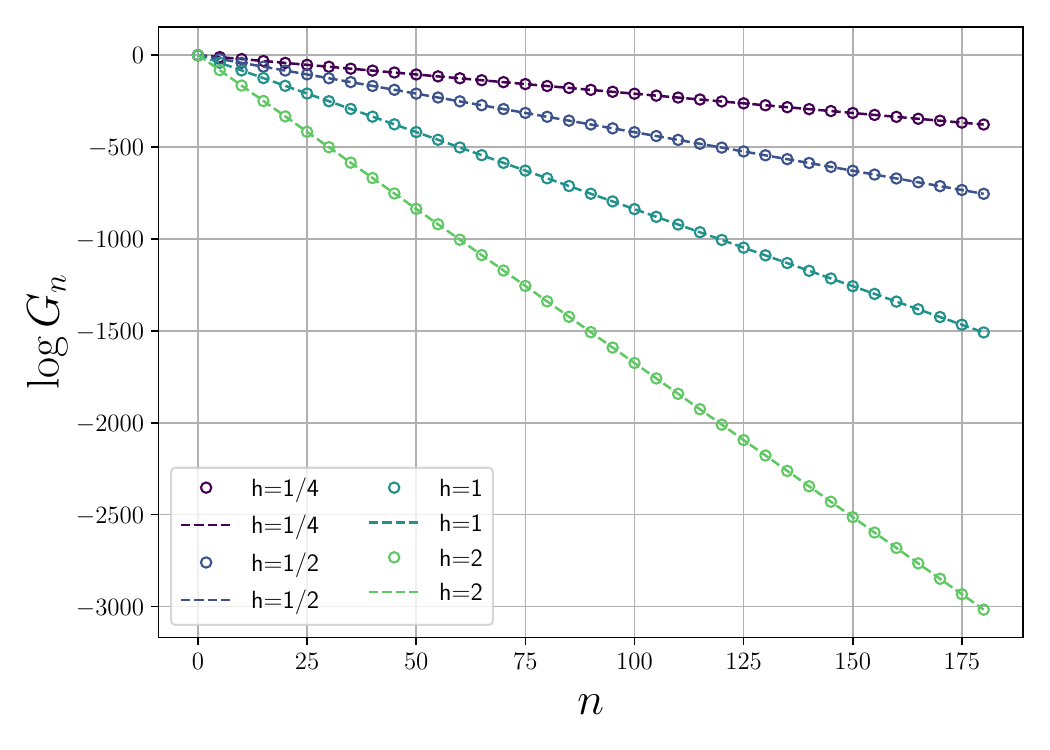}
    \caption{$(T_0/L, T_1/L) = (1.8, 1.8)$, $\eta > 1$}
    \label{fig:heating Gn 2}
  \end{subfigure}
  \caption{$\log G_n$ vs $n$ in the heating phase, with $\eta < 1$ in (a) and $\eta > 1$ in (b). The dots show the values obtained from the exact high-precision CFT computations, while the dashed lines show the approximation in \eqref{eq:Gn heating eta both}, for $h = 1/4, 1/2, 1, 2$. }
  \label{fig:Gn heating}
\end{figure}

In the heating phase, one can show:
\begin{align}
	\eta = 
	\begin{cases}
		e^{-\varphi},  ~~ \text{for, } (2m+1)L > T_0 > 2mL$, $m\in\mathbb Z_{\ge0}, \\
		e^{\varphi'} ~~ \text{for, } (2m+2)L > T_0 > (2m+1)L$, $m\in\mathbb Z_{\ge0},
	\end{cases}
\end{align}
with $\varphi, \varphi' > 0$. Following this, one can then approximate $G_n$ from  \eqref{eq:Gn}:
\begin{equation}  \label{eq:Gn heating eta both}
	G_n \simeq 
	\begin{cases}
		\displaystyle \left| \frac{\gamma _1-\gamma _2}{\gamma _2} \right|^{4 h} |\eta|^{2 n h} & \text{for } (2m+1)L > T_0 > 2mL,\\[10pt]
		\displaystyle \left| \frac{\gamma _1-\gamma _2}{\gamma _1} \right|^{4 h} |\eta|^{-2 n h} & \text{for } (2m+2)L > T_0 > (2m+1)L.
	\end{cases}
\end{equation}
A detailed derivation of the above is discussed in $\S$\ref{app:Gn}. In Fig.\ref{fig:Gn heating}, we plot the $G_n$ values obtained from the exact CFT computations in both the regimes in the non-heating phase, and compare them with the approximate expression in \eqref{eq:Gn heating eta both}. 

\begin{figure}[htbp]
  \centering
  \begin{subfigure}{0.45\textwidth}
    \centering
    \includegraphics[width=\linewidth]{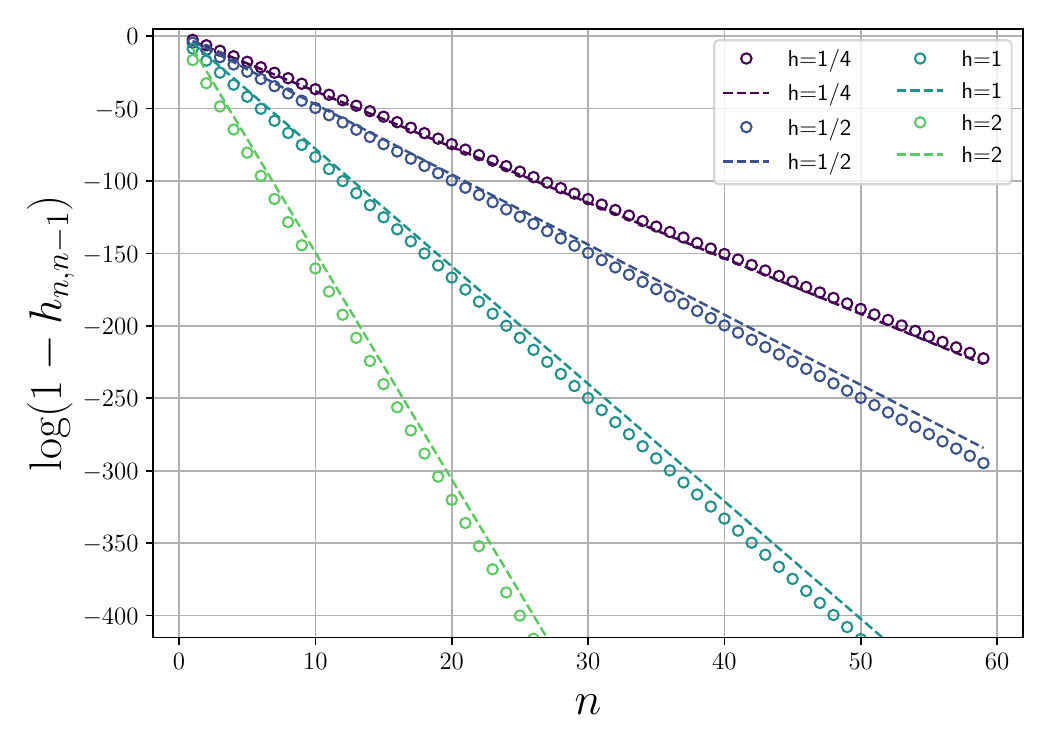}
    \caption{$(T_0/L, T_1/L) = (0.8, 0.8)$, $\eta < 1$}
    \label{fig:heating lanczos 1}
  \end{subfigure}\hfill
  \begin{subfigure}{0.45\textwidth}
    \centering
    \includegraphics[width=\linewidth]{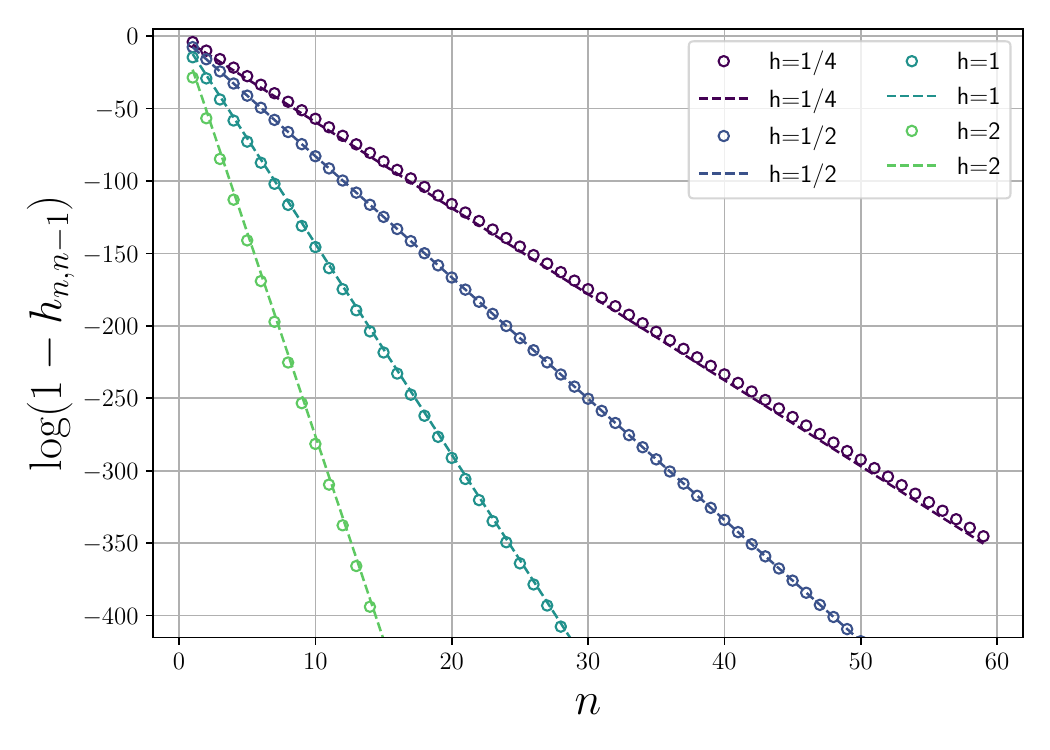}
    \caption{$(T_0/L, T_1/L) = (1.8, 1.8)$, $\eta > 1$}
    \label{fig:heating lanczos 2}
  \end{subfigure}
  \caption{$h_{n,n-1}$ as function of $n$ in the heating phase of square wave drive. The dots denote the exact coefficients, while the dashed lines denote the approximation shown in \eqref{heating lanczos ssd approx}. }
  \label{fig:lanczos heating}
\end{figure}

With these approximations, it is possible to calculate the Arnoldi coefficients $h_{n,n-1}$ analytically. For example, the first two Arnoldi coefficients are:
\begin{align}
    h_{1,0} = \sqrt{1-G_1^2}, ~~ h_{2,1} = \sqrt{\frac{\left(2 G_1^2-G_2-1\right) \left(G_2-1\right)}{\left(G_1^2-1\right){}^2}}.
\end{align}
Since $\varphi, \varphi'$ are positive real quantities, it is possible to do an expansion in terms of $e^{-\varphi} \, (e^{-\varphi'})$. In the leading order, one can express the Arnoldi coefficients as the following,
\begin{align}
    h_{n,n-1} \simeq 
    \begin{cases}
        1 - \frac{1}{2} p^2 (p-1)^{2 n}  e^{-4 n h \varphi} ~~  \text{for } (2m+1) L > T_0 > 2m L, \\[8pt]
        1 - \frac{1}{2} p'^2 (p'-1)^{2 n}  e^{-4 n h \varphi'} ~~  \text{for } (2m+2) L > T_0 > (2m+1) L, 
    \end{cases}
\end{align}
where,
\begin{align}
    p = \left| \dfrac{\gamma _1-\gamma _2}{\gamma _2} \right|^{4 h} , ~~ p' = \left| \dfrac{\gamma _1-\gamma _2}{\gamma _1} \right|^{4 h}.
\end{align}
In terms of $\eta$, the Arnoldi coefficients behave as,
\begin{align} \label{heating lanczos ssd approx}
    h_{n,n-1} \simeq 
    \begin{cases}
        1 - \frac{1}{2} p^2 (p-1)^{2 n} \, \eta^{4 k h} ~~  \text{for } (2m+1) L > T_0 > 2m L, \\[8pt]
        1 - \frac{1}{2} p'^2 (p'-1)^{2 n} \, \eta^{- 4 k h} ~~  \text{for } (2m+2) L > T_0 > (2m+1) L. 
    \end{cases}
\end{align}
\noindent In Fig.\ref{fig:lanczos heating}, the behaviour of the exact Arnoldi coefficients in the haeting phase are shown, along with the approximation obtained in \eqref{heating lanczos ssd approx}.

\subsubsection{The non-heating phase}

\begin{figure}[htbp]
  \centering
  \begin{subfigure}{0.45\textwidth}
    \centering
    \includegraphics[width=\linewidth]{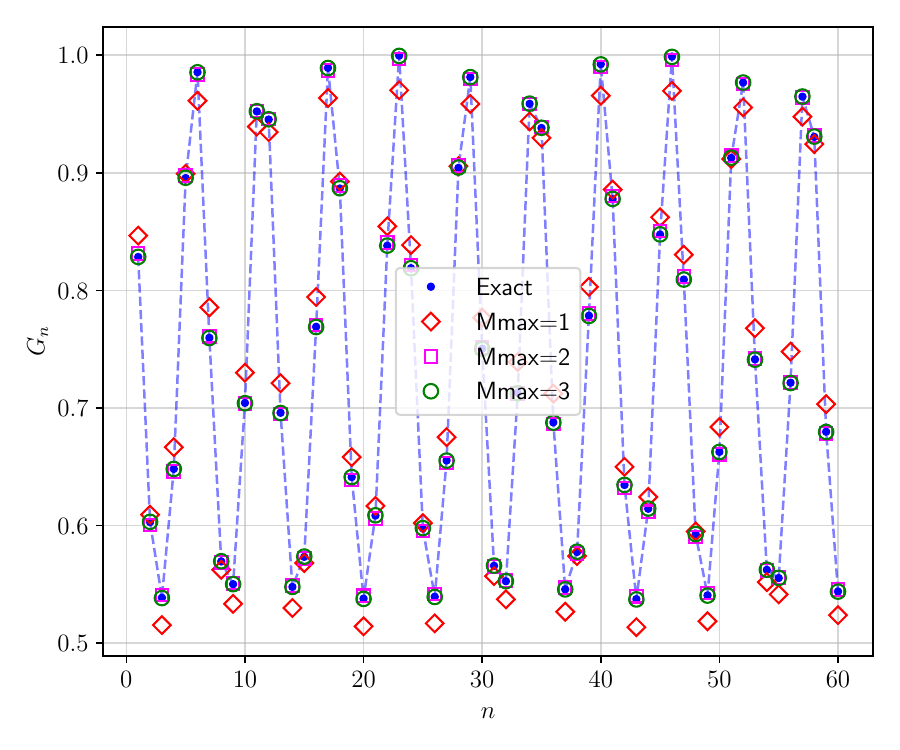}
    \caption{$(T_0/L, T_1/L) = (0.1, 0.1)$}
    \label{fig:non-heating Gn 1}
  \end{subfigure}\hfill
  \begin{subfigure}{0.45\textwidth}
    \centering
    \includegraphics[width=\linewidth]{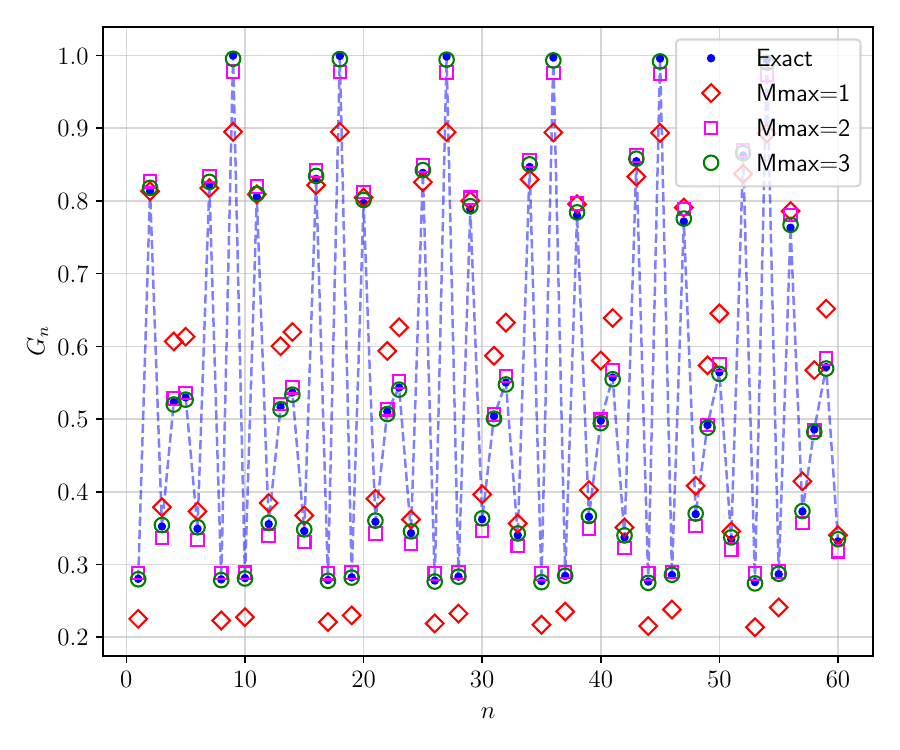}
    \caption{$(T_0/L, T_1/L) = (0.3, 0.3)$}
    \label{fig:non-heating Gn 2}
  \end{subfigure}
  \caption{$G_n$ vs $n$ in non-heating phase, comparison between exact values and analytical approximations. $Mmax$ denotes the number of terms in the sum in \eqref{eq:Gn non-heating in r}. }
  \label{fig:Gn non-heating}
\end{figure}

In the non-heating phase, one has $\eta = e^{i \phi}$ with $\phi \in \mathbb{R}$, and the corresponding expression of $G_n$ becomes:
\begin{equation} \label{eq:Gn non-heating in r}
	\begin{aligned}
		G_n
		&=
		(1-2r+r^2)^{2h}
		\Bigg\{
		{}_2F_1(2h,2h;1;r^2)
		\\
		&\qquad
		+
		2\sum_{m=1}^{\infty}
		r^{m}
		\binom{2h+m-1}{m}
		{}_2F_1\!\big(2h,2h+m;m+1;r^2\big)
		\cos(m n\phi)
		\Bigg\},
	\end{aligned}
\end{equation}
where,
\begin{equation}
	r=
	\begin{cases}
		\dfrac{|\gamma_2|}{|\gamma_1|}  ~~~ \text{for } |\gamma_1| > |\gamma_2|, \, \text{i.e., } \sin\frac{\pi T_0}{L} + \frac{\pi T_1}{L} \cos\frac{\pi T_0}{L} < 0, \\[10pt]
		\dfrac{|\gamma_1|}{|\gamma_2|}  ~~~ \text{for } |\gamma_2| > |\gamma_1|, \, \text{i.e., } \sin\frac{\pi T_0}{L} + \frac{\pi T_1}{L} \cos\frac{\pi T_0}{L} > 0.
	\end{cases}
\end{equation}
See $\S$\ref{app:Gn nonheating} for more details. In the non-heating phase, the series above indeed converges quickly for very few terms. For example, in Fig.\ref{fig:Gn non-heating} we compare the exact $G_n$ with the truncated sum in \eqref{eq:Gn non-heating in r}, which matches quite well. 

\begin{figure}[htbp]
    \centering
    \includegraphics[width=0.45\textwidth]{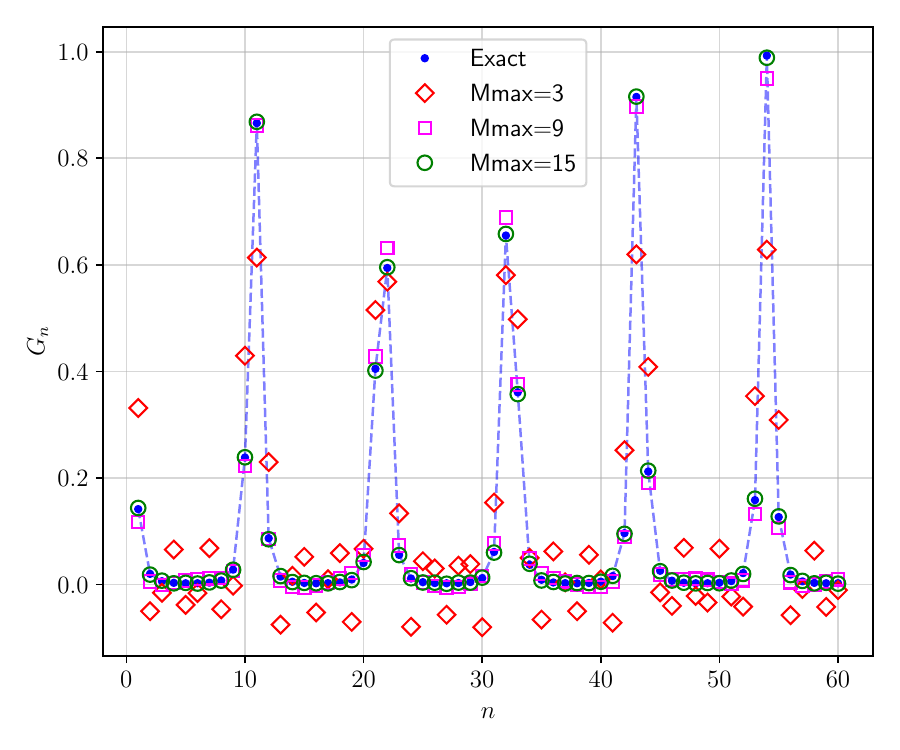}
    \caption{$G_n$ vs $n$ at a point (here, $(T_0/L, T_1/L) = (0.41, 0.41)$) close to the critical line, approaching from the non-heating phase.}
    \label{fig:critical Gn}
\end{figure}

On the other hand, as one gets closer to the critical point, the expression in \eqref{eq:Gn non-heating in r} starts to converge slowly. This is expected since the sum is obtained from the series expansion in terms of the parameter $r$, which gets closer to 1 (as $|\gamma_1| \simeq |\gamma_2|$) as one approaches the critical point. As an example, we see in Fig.\ref{fig:critical Gn} that one needs to keep more terms in expression \eqref{eq:Gn non-heating in r} to closely approximate the values obtained from exact CFT computations.

\section{Square wave drive in critical fermions}
\label{sec-lat}

We consider a one-dimensional lattice of spinless fermions governed by the Hamiltonians \cite{PhysRevB.97.184309,Wen:2018agb,Katsura:2011ss,Katsura:2011zyx,PhysRevB.84.165132}
\begin{equation}
H^{\mathrm{lat}}_0
= \frac{1}{2}\sum_{i=1}^{L-1}\!\left( c_i^\dagger c_{i+1} + \mathrm{h.c.} \right),
\qquad
H^{\mathrm{lat}}_1
= \sum_{i=1}^{L-1}
\sin^2\!\left(\frac{\pi\,(i+\tfrac{1}{2})}{L}\right)
\left( c_i^\dagger c_{i+1} + \mathrm{h.c.} \right),
\label{eq:Hlat}
\end{equation}
where \(c_i\) and \(c_i^\dagger\) are fermionic annihilation and creation operators satisfying
\(
\{c_i,c_j\}=\{c_i^\dagger,c_j^\dagger\}=0
\)
and
\(
\{c_i,c_j^\dagger\}=\delta_{ij}.
\)

At time \(t=0\) the system is prepared in the ground state \(|G\rangle\) of \(H^{\mathrm{lat}}_0\), corresponding to the half-filled Fermi sea of the uniform hopping model. The system is then evolved under \(H^{\mathrm{lat}}_1\) for a time \(T_1\), followed by evolution under \(H^{\mathrm{lat}}_0\) for a time \(T_0\). This two-step driving protocol is repeated periodically, defining the Floquet dynamics analogous to the Floquet CFT setup discussed in the previous section.

\subsection{Krylov construction with open boundary conditions}

We characterize the Floquet dynamics through the many-body return amplitude
\begin{equation}
G_m = \langle G|U_F^{\,m}|G\rangle ,
\end{equation}
which measures the overlap between the initial state and the state evolved for \(m\) Floquet cycles.

Let \(\{\varepsilon_\alpha,\ket{\varphi_\alpha}\}\) denote the eigenpairs of \(H^{\mathrm{lat}}_0\) in the single particle sector,
\begin{equation}
H^{\mathrm{lat}}_0 \ket{\varphi_\alpha} = \varepsilon_\alpha \ket{\varphi_\alpha},
\qquad \alpha=1,\dots,L,
\end{equation}
with eigenvalues ordered increasingly. The many-body ground state corresponds to the half-filled Fermi sea obtained by occupying the lowest \(L/2\) single-particle modes. We collect these modes into the matrix
\begin{equation}
\Phi = \bigl( \ket{\varphi_1}, \ket{\varphi_2}, \dots, \ket{\varphi_{L/2}} \bigr).
\end{equation}

The single-particle Floquet operator for one driving period is
\begin{equation}
U_F = e^{-i H^{\mathrm{lat}}_0 T_0}\, e^{-i H^{\mathrm{lat}}_1 T_1}.
\end{equation}

Due to the free-fermion structure, the return amplitude can be written as ~\cite{2003JPhA...36L.205P,Adhikari:2023evu,Xia:2024xfd}
\begin{equation}
G_m
= \det\!\left(\Phi^\dagger U_F^{\,m}\Phi\right).
\label{eq:return_lat}
\end{equation}

In the low-energy regime the lattice dynamics admits an effective conformal field theory description. In the CFT picture the Floquet evolution corresponds to a Möbius transformation
\begin{equation}
z \mapsto z_m = \frac{a_m z + b_m}{c_m z + d_m}.
\end{equation}

The CFT prediction for the return amplitude takes the universal form
\begin{equation}
G_m^{\mathrm{CFT}}
=
\left|
\frac{\partial z_m}{\partial z}\Big|_{z=0}
\right|^{h},
\label{eq:return_cft}
\end{equation}
where \(h=\frac{1}{16}\) is the scaling dimension of the boundary operator associated with the initial state \cite{Konechny:2018ujl}.

\begin{figure}[t]
\centering
\begin{subfigure}{0.45\textwidth}
\centering
\includegraphics[width=\linewidth]{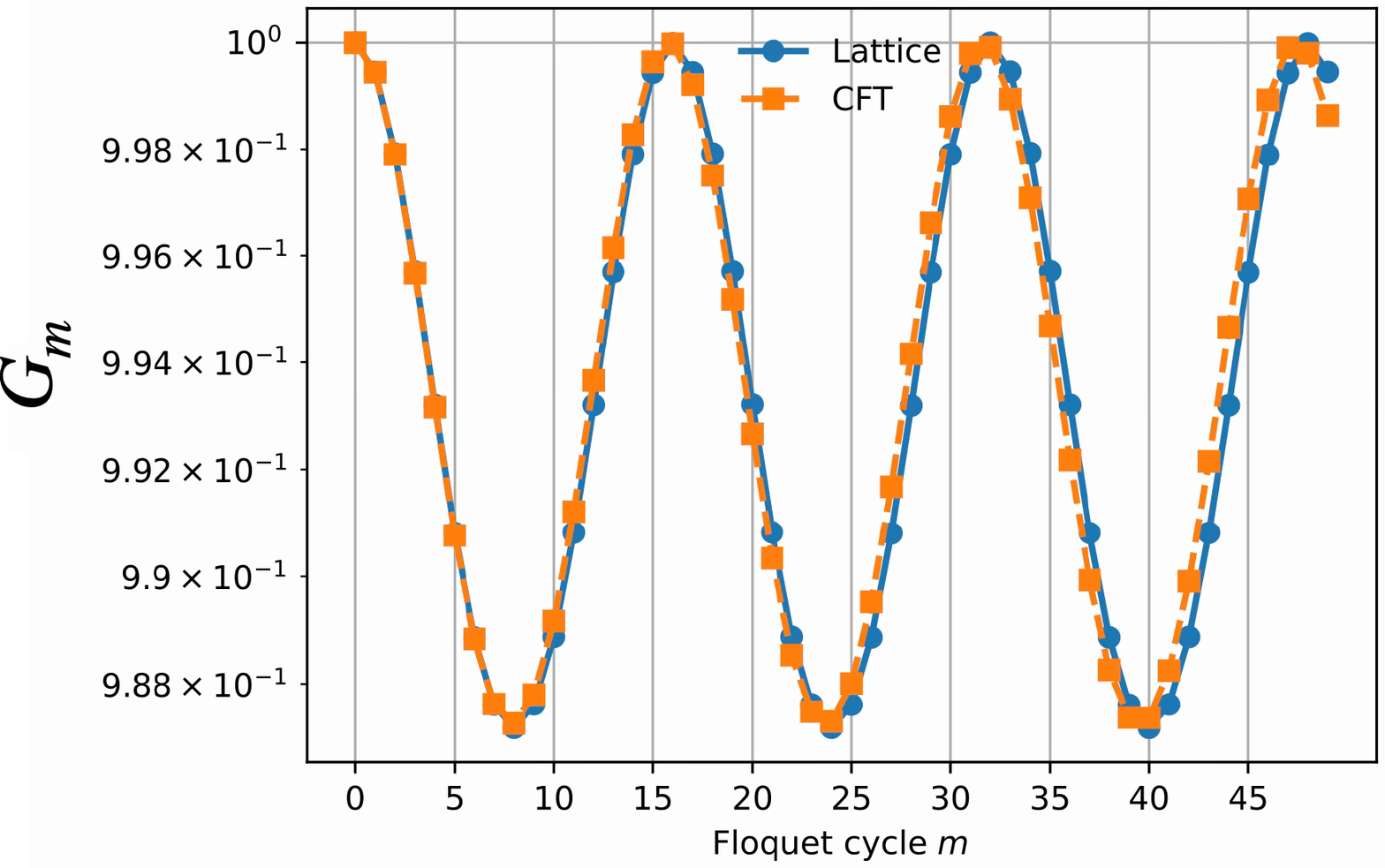}
\caption{Non-heating phase \((T_0/L,T_1/L)=(0.04,0.03)\).}
\end{subfigure}
\hfill
\begin{subfigure}{0.45\textwidth}
\centering
\includegraphics[width=\linewidth]{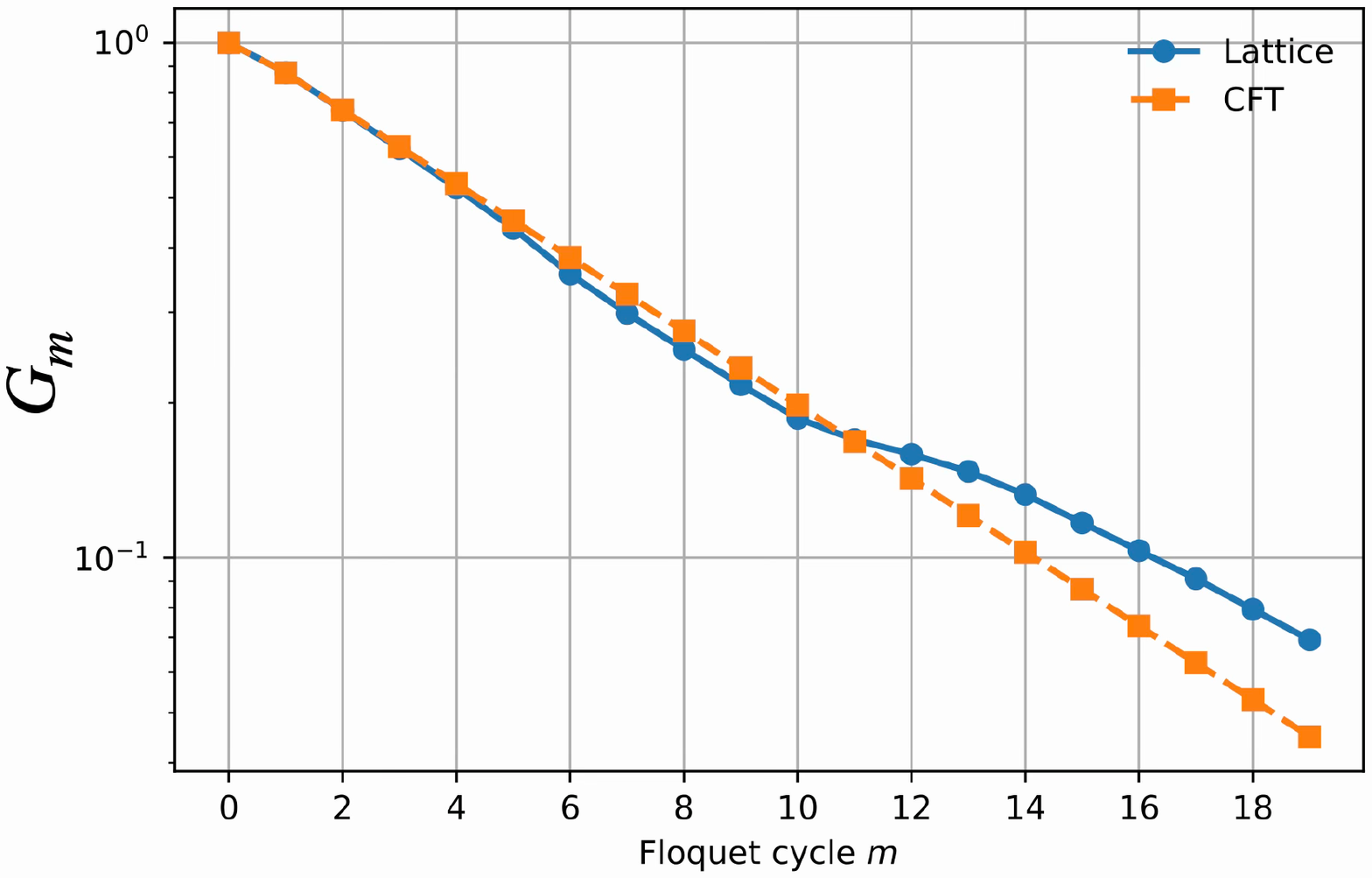}
\caption{Heating phase \((T_0/L,T_1/L)=(0.87,0.87)\).}
\end{subfigure}
\caption{Return amplitude \(G_m\) in different dynamical regimes of the Floquet drive.}
\label{fig:return_amp_phases}
\end{figure}

Figure~\ref{fig:return_amp_phases} compares the lattice  of the return amplitude $G_m$ ($L =128$ ) with the CFT prediction and demonstrates quantitative agreement over many Floquet cycles.

\begin{figure}
\centering
\begin{subfigure}{0.48\textwidth}
\centering
\includegraphics[width=\linewidth]{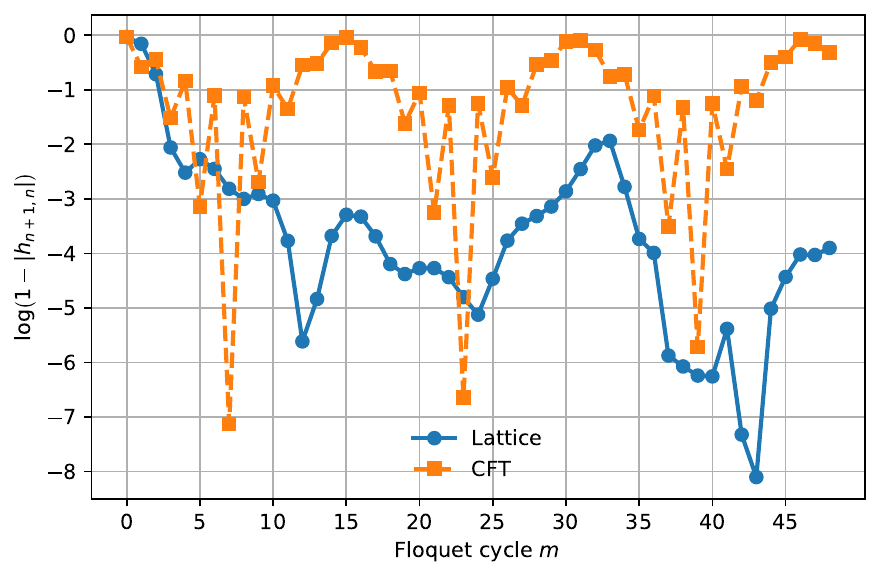}
\caption{Non-heating phase.}
\end{subfigure}
\hfill
\begin{subfigure}{0.48\textwidth}
\centering
\includegraphics[width=\linewidth]{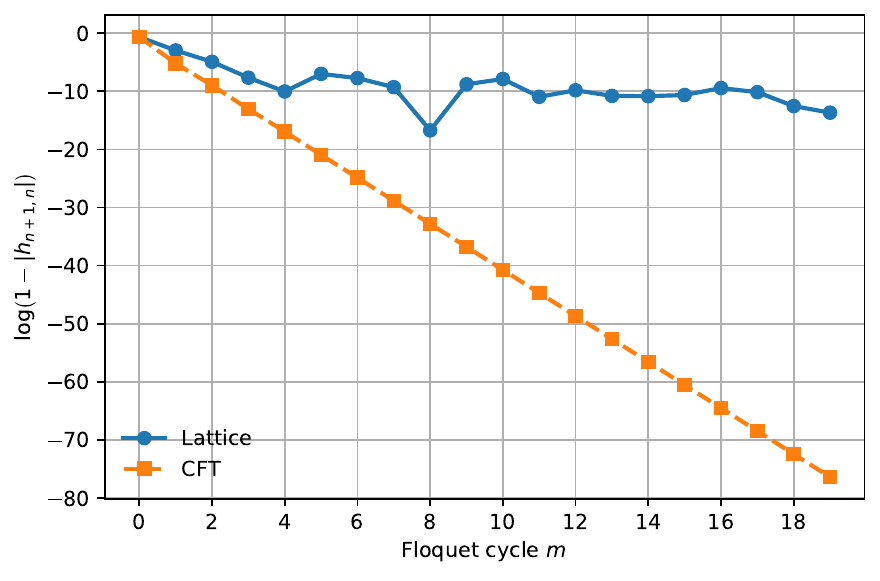}
\caption{Heating phase.}
\end{subfigure}
\caption{Sub-diagonal matrix elements \(h_{n,n-1}\) in different dynamical regimes.}
\label{fig:log_dev_sl2}
\end{figure}

We next examine Fig.~\ref{fig:log_dev_sl2} the structure of the sub-diagonal matrix elements \(h_{n,n-1}\) of the Floquet unitary \(U_F\) in the Krylov basis constructed from the initial state ($L =128$ ). These coefficients characterize the operator growth generated by the Floquet dynamics.

In the non-heating phase the coefficients \(h_{n,n-1}\) increase gradually with irregular fluctuations. In contrast, in the heating phase they rapidly approach unity and subsequently saturate. The non-fluctuating behavior of \(h_{n,n-1}\) may provide a useful diagnostic distinguishing heating from non-heating dynamics.

\subsection{K-complexity of the correlation matrix.}

In this subsection, we construct the Krylov basis using the time evolution of the correlation matrix in the single particle sector~\cite{Caputa:2024vrn}. This approach provides direct access to the full set of Arnoldi coefficients $h^{C}_{n+1,n}$ without truncation. Moreover, it enables the computation of the Krylov complexity $C_K(j)$ to arbitrarily late times, allowing us to probe the long-time operator dynamics in a controlled and efficient manner for finite size systems.

\paragraph{Relation to the Fock space description:}

We defined $\Phi \in \mathbb{C}^{N\times M}$ ($M=N/2$) to denote the matrix of occupied single-particle modes, so that the initial correlation matrix is
\[
C(0)=\Phi \Phi^\dagger.
\]
Define the overlap matrix
\[
O_m = \Phi^\dagger U_F^m \Phi.
\]

The many-body overlap between the initial and evolved Slater determinants is given by~\cite{2003JPhA...36L.205P}
\[
G_{m} = \langle \Psi_0 | U_F^m | \Psi_0 \rangle
= \det(O_m).
\]
If $\{\lambda_k\}$ are the eigenvalues of $O_m$, this becomes
\[
G_{m} = \prod_{k=1}^{M} \lambda_k.
\]

On the other hand, the autocorrelation of $C(0)$ in single particle space is
\[
g_{m} = \frac{\mathrm{Tr}\big(C(0)\,C(m)\big)}{\mathrm{Tr}\big(C(0)^2\big)},
\qquad
C(m)=U_F^m C(0) U_F^{\dagger m}.
\]
Using $C(0)=\Phi\Phi^\dagger$ and cyclicity of the trace,
\[
\mathrm{Tr}\big(C(0)C(m)\big)
= \mathrm{Tr}\big(O_m O_m^\dagger\big).
\]
Hence,
\[
g_{m} = \frac{1}{\mathrm{Tr}\big(C(0)^2\big)}\,\mathrm{Tr}\big(O_m O_m^\dagger\big)
= \frac{1}{\mathrm{Tr}\big(C(0)^2\big)}\sum_{k=1}^{M} \sigma_k^2,
\]

\noindent where $\sigma_k$ are the singular values of $O_m$. Thus, both $G_{m}$ and $g_{m}$ are determined by the same overlap matrix $O_m$ as \( |G_{m}|^2 = \prod_{k=1}^{M} \sigma_k^2\). As explained in Sec.~2, the Krylov basis can be constructed solely from the return amplitudes. Therefore, the correspondence between $G(m)$ and $g(m)$ can be used to relate the Arnoldi coefficients which are generated from the many body overlap ($h_{n+1.n}$) and autocorrelation of correlation  matrix($h^{C}_{n+1,n}$).

\begin{figure}[t]
\centering

\includegraphics[width=0.48\linewidth]{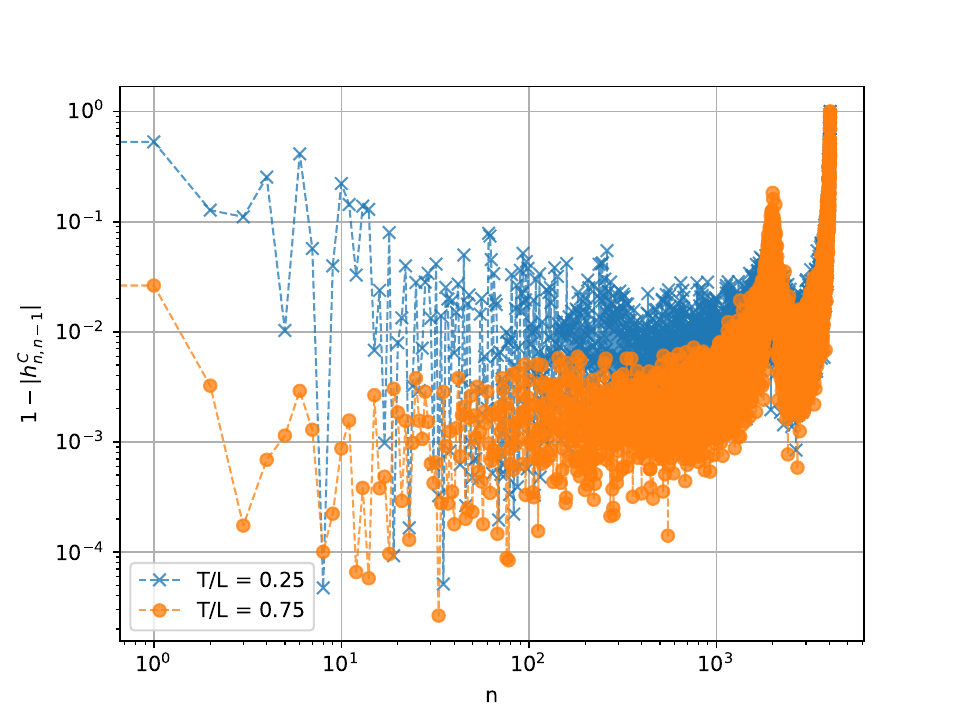}
\hfill
\includegraphics[width=0.48\linewidth]{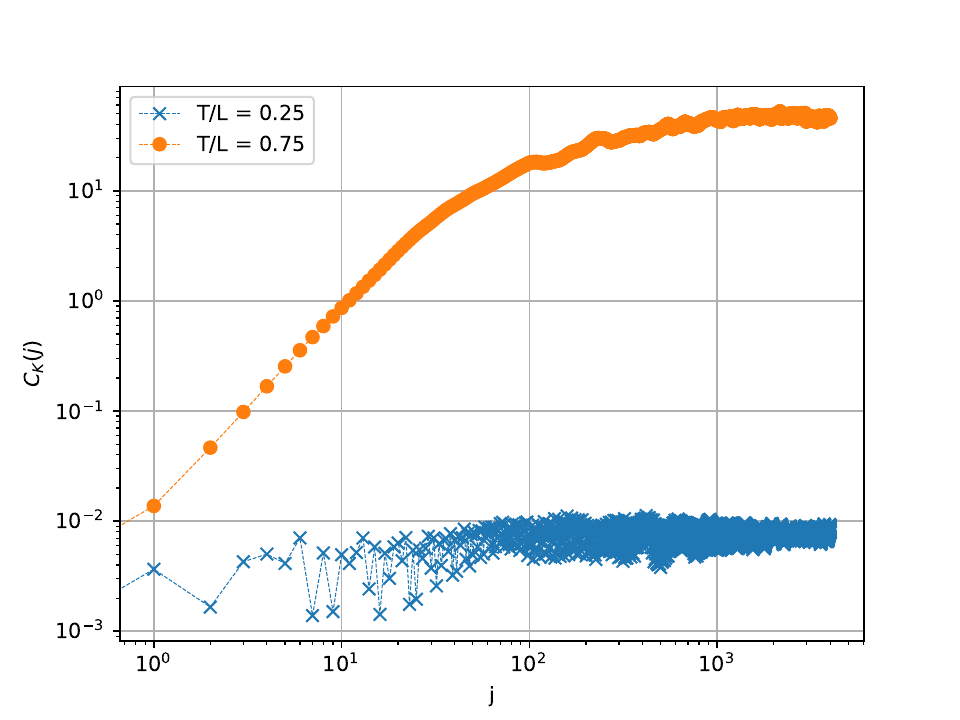}

\caption{ (left) sub-diagonal matrix elements and K -complexity (right) for the Correlation Matrix. The blue and orange curves denote the behaviour in the non-heating and the heating phase, respectively. }
 
\label{fig:full_ssd_krylov}
\end{figure}

For $L = 64$, we construct the full Krylov basis, enabling access to all sub-diagonal coefficients $h^{C}_{n+1,n}$ and the corresponding Krylov complexity $C_K(j)$, as shown in Fig.\ref{fig:full_ssd_krylov}. Under $j$ Floquet period, the correlation matrix evolves unitarily as
\begin{equation}
C(j)=U_F^{\,j}\,C(0)\,U_F^{\dagger\, j},
\end{equation}

Now, for this time evolution, one can construct the Krylov basis using Eq.~\ref{kconst}, with the initial condition
\[
|K_0) = \frac{|C(0))}{\|C(0)\|}.
\]

This gives the Krylov complexity of the correlation matrix as~\cite{Caputa:2024vrn}
\[
C_K(j) = \sum_{n} n \, \big| (K_n \,|\, \mathcal{U}_F^{\,j} \,|\, K_0) \big|^2. 
\]

\noindent where $\mathcal U^{j}|X) \equiv U_F^{j} X U_F^{\dagger j} $ and the inner product is defined as the Hilbert--Schmidt inner product,\(\
(A|B) = \mathrm{Tr}\!\left(A^\dagger B\right)\). For each choice $\tfrac{T_0}{L}=\tfrac{T_1}{L}=\tfrac{T}{L}$ in $U_F$, we sample an ensemble of $n_{\mathrm{samples}}=10$ realizations with $\sim 5\%$ uniform fluctuations around the base values (scaled with system size), and report ensemble-averaged results in Fig.\ref{fig:full_ssd_krylov}. In the heating regime, the sub-diagonal coefficients initially rise rapidly and approach unity, similar to the Lanczos ascent \cite{Kar:2021nbm} in time-independent chaotic Hamiltonian evolution. At intermediate Krylov steps, they exhibit oscillations around a nearly constant value, forming the so-called Lanczos plateau \cite{Kar:2021nbm}. At later steps, the coefficients decay toward zero, reflecting the eventual finite-size saturation of the Krylov chain. In the non-heating regime, the initial growth of the sub-diagonal coefficients is suppressed. The resulting Lanczos plateau forms at values significantly below unity, and at later Krylov steps the coefficients again decay to zero. The early growth of the sub-diagonal coefficients reflects scrambling in single-particle mode space and persists until the initially local operator becomes mixed across momentum modes. 

The K-complexity exhibits markedly different behaviour in the two phases at early Floquet steps. In the heating phase, it shows a prominent growth followed by saturation at late times. In contrast, during the non-heating phase, there is no sustained growth; instead, the complexity oscillates around a small mean value, which remains significantly below the plateau reached in the heating phase.

\paragraph{Krylov construction with periodic boundary conditions:}

We now repeat the analysis for periodic boundary conditions. In this case the same Hamiltonians \eqref{eq:Hlat} are used but with hopping terms connecting the first and last sites. The initial state is chosen to be the Fermi sea together with the two lowest-lying excitations~\cite{Berganza:2011mh}. This setup allows the study of states with different conformal dimension \(h = \frac{1}{2}\).

\begin{figure}[t]
\centering
\begin{subfigure}{0.45\textwidth}
\centering
\includegraphics[width=\linewidth]{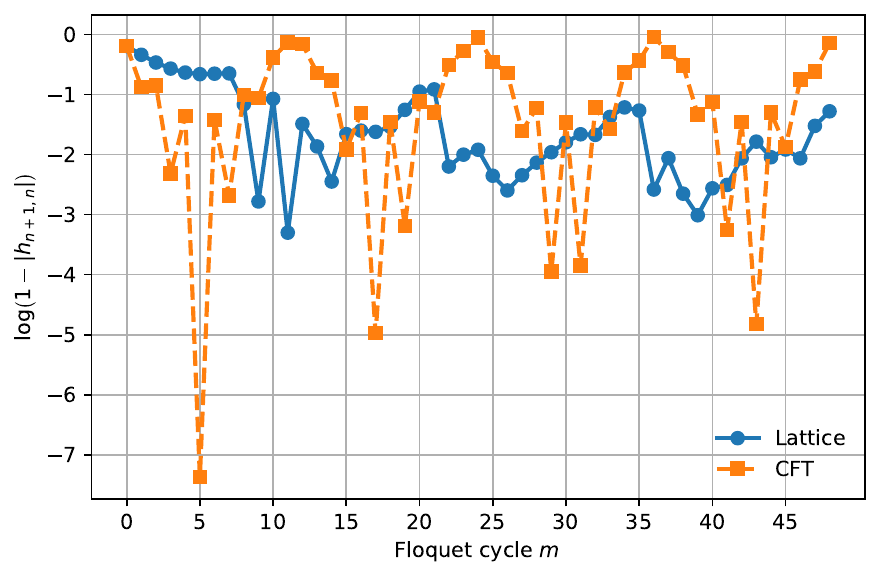}
\caption{Non-heating phase.}
\end{subfigure}
\hfill
\begin{subfigure}{0.45\textwidth}
\centering
\includegraphics[width=\linewidth]{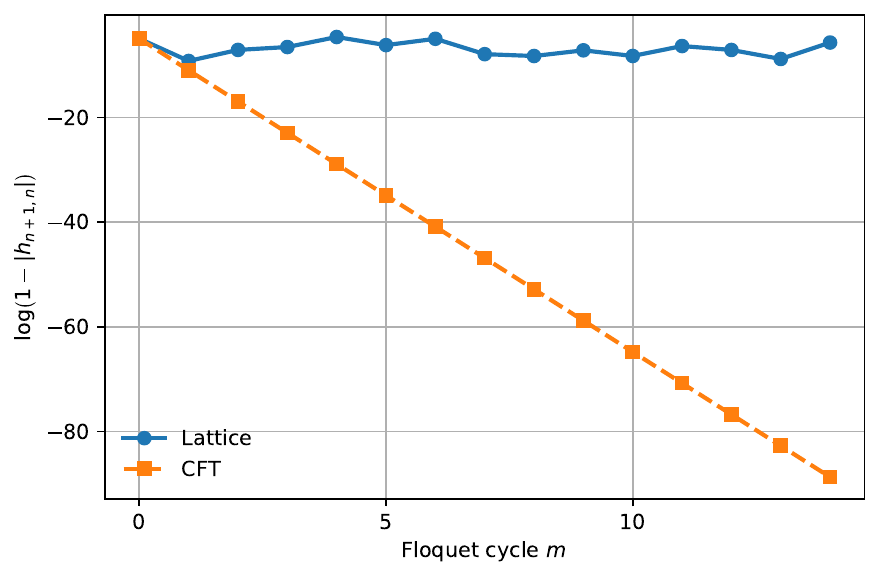}
\caption{Heating phase.}
\end{subfigure}
\caption{sub-diagonal Krylov matrix elements with periodic boundary conditions.}
\label{fig:p_log_dev_sl2}
\end{figure}

Figure~\ref{fig:p_log_dev_sl2} shows the behaviour of the Arnoldi coefficients for the state with conformal dimension \(h=\tfrac12\). The qualitative behaviour remains similar to the open boundary case: in the non-heating phase the coefficients show fluctuations, whereas in the heating phase they saturate rapidly.

\subsection{Spectral Analysis and late-time return amplitude}

To further characterize the dynamical regimes we analyze the spectral statistics of the Floquet operator. The quasienergy spectrum is obtained from the eigenvalues \(e^{i\theta_\alpha}\) of \(U_F\), with quasienergies \(\varepsilon_\alpha=-\theta_\alpha\) mapped to the interval \((-\pi,\pi]\).

We compute the ratio of adjacent quasienergy spacings
\begin{equation}
r_\alpha = \frac{\min(s_\alpha,s_{\alpha+1})}{\max(s_\alpha,s_{\alpha+1})},
\end{equation}
where \(s_\alpha=\varepsilon_{\alpha+1}-\varepsilon_\alpha\). The mean value \(\langle r\rangle\) distinguishes Poisson statistics (\(\approx0.386\)) from Wigner--Dyson statistics (\(\approx0.53\)) belonging to Circular Orthogonal ensemble.

\begin{figure}[t]
\centering
\begin{subfigure}{0.45\textwidth}
\includegraphics[width=\linewidth]{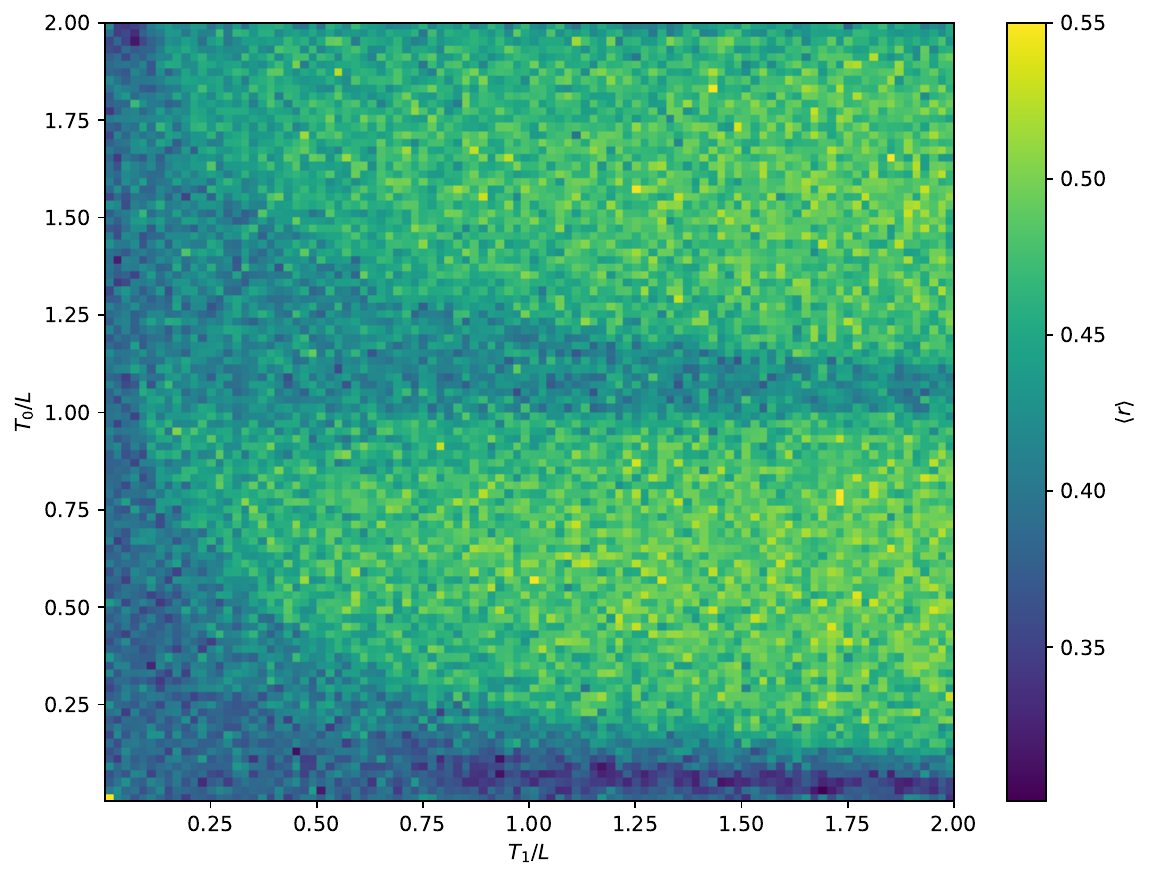}
\caption{\(r\)- factor phase diagram.}
\end{subfigure}
\hfill
\begin{subfigure}{0.45\textwidth}
\includegraphics[width=\linewidth]{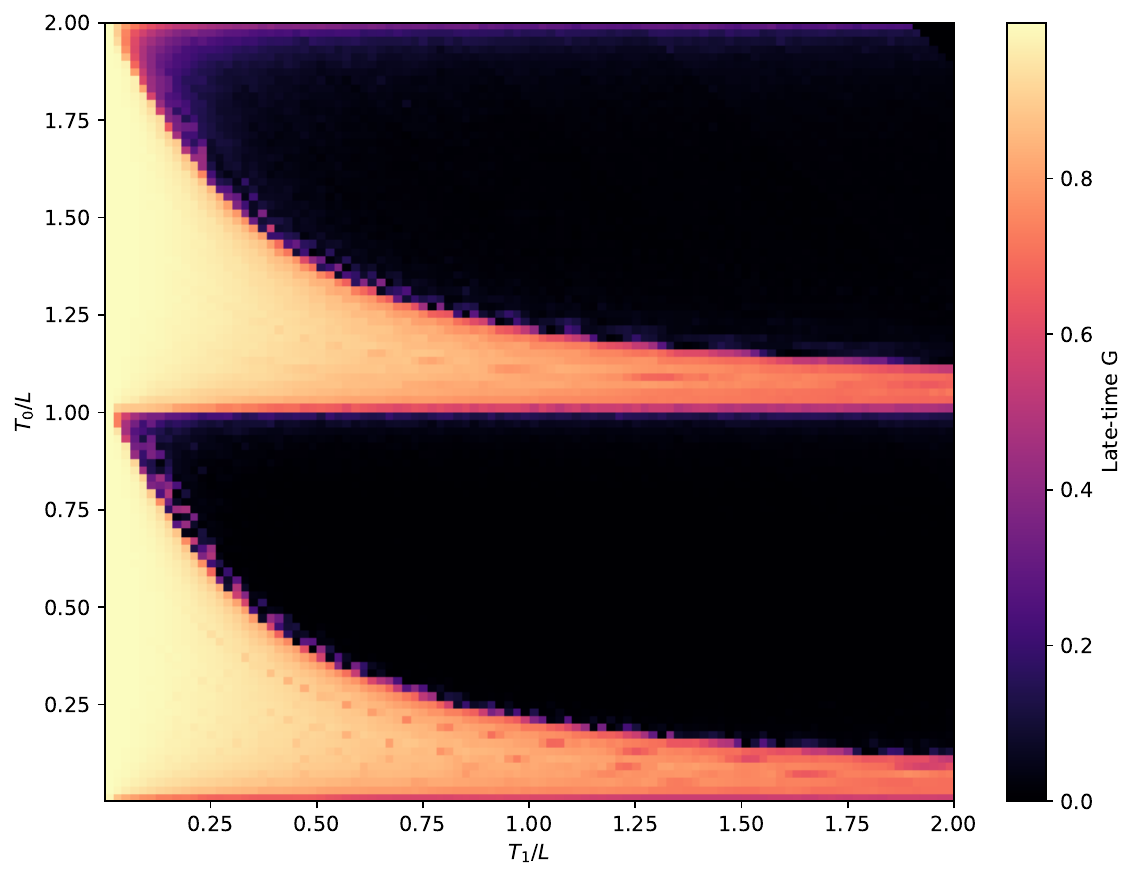}
\caption{Late-time Loschmidt echo.}
\end{subfigure}
\caption{Floquet phase diagram in the \((T_0/L,T_1/L)\) plane obtained from spectral statistics and the late-time Loschmidt echo.}
\label{fig:phas}
\end{figure}

Figure~\ref{fig:phas} shows the resulting phase diagram together with the late-time Loschmidt echo. In the heating phase the spectrum follows Wigner--Dyson statistics and the return amplitude decays rapidly, leading to a vanishing Loschmidt echo at late times. In contrast, in the non-heating phase the spectrum exhibits Poisson-like statistics and the Loschmidt echo displays persistent oscillatory behaviour.

\begin{figure}[t]
\centering
\includegraphics[width=0.45\linewidth]{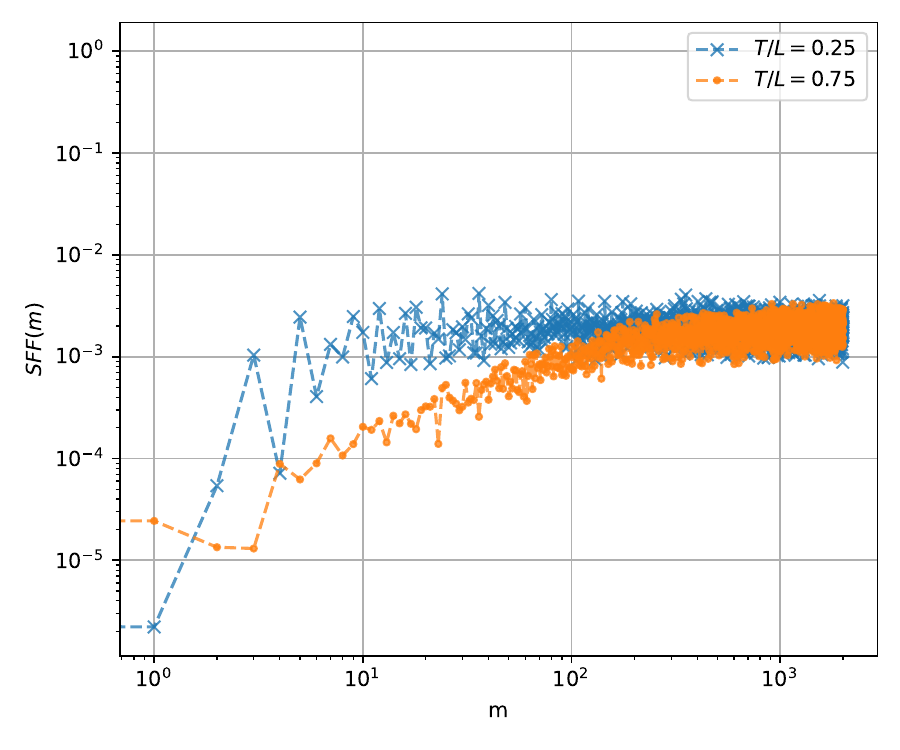}

\caption{
Spectral form factor $\mathrm{SFF}(m)$ for $N=512$ shown on a log--log scale for two driving parameters, $T/L = 0.25$ (non-heating) and $T/L = 0.75$ (heating).
}
\label{fig:ssd_sff}
\end{figure}

The spectral form factor (SFF) is also computed for the Floquet unitary $U_F$~\cite{2001JPhA...34.8485T} as
\[
\mathrm{SFF}(m) = \frac{1}{N^2}\,\big|\mathrm{Tr}(U_F^m)\big|^2,
\]
which probes correlations in the quasienergy spectrum of the unitary dynamics. For each choice of driving parameters, the SFF is evaluated and averaged over an ensemble of 40 realizations. The ensemble is generated by introducing small fluctuations in the driving parameters around their base values, ensuring a controlled averaging over nearby unitaries. From Figure~\ref{fig:ssd_sff} in the heating phase, the quasienergy level spacing distribution follows Wigner--Dyson statistics, reflecting chaotic spectral correlations. Correspondingly, the SFF exhibits the characteristic dip--ramp--plateau structure, with a linear ramp followed by saturation at late times. In contrast, in the non-heating phase, the spectrum deviates from Wigner--Dyson behavior, and the SFF shows no clear linear ramp; instead the SFF sharply rises directly approaching a plateau, indicating the absence of strong level repulsion. This sharp rise of SFF resembles the behaviour observed in disordered two-body systems, such as SYK-2 \cite{Winer:2020mdc} and disorder averaged topological insulators \cite{Sarkar:2023pdf}.

\begin{figure}[t]
\centering
\includegraphics[width=0.45\textwidth]{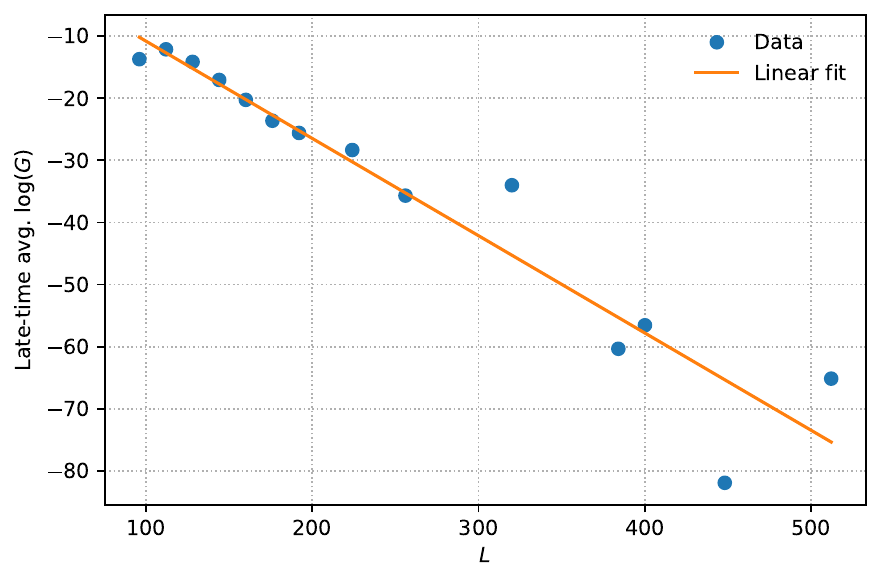}
\caption{System-size scaling of the late-time Loschmidt echo.}
\label{fig:ls_scale}
\end{figure}

Finally, we study the scaling of the late-time Loschmidt echo with system size in the heating phase \ref{fig:ls_scale}. The system is evolved with equal driving times \(T_0=T_1=T\) at fixed \(T/L=0.75\).

The late-time value of \(\log \mathcal{L}\) decreases approximately linearly with system size \(L\), indicating an exponential suppression
\begin{equation}
\mathcal{L}\sim e^{-cL}.
\end{equation}

This behaviour is consistent with heating dynamics, where the return amplitude vanishes in the thermodynamic limit.

\section{Continuous drive in CFT}

For CFTs with a sine-square deformations, one can write down a generalized expression of the driven system which does not break the conformal symmetry of the system. Here the driven Hamiltonian takes the form \cite{Das:2021gts},
\begin{align}
    H(t) = \frac{2\pi}{L} \left[ f(t) L_0 + \frac{1}{2} f_1(t) \left( L_1 + L_{-1} \right) \right] .
\end{align}
Note, for the drive discussed in the earlier sections, one has $f(t) = 1$ and $f_1(t)$ is a square wave pulse:
\begin{align*}
f_1(t)=\sum_{n\in\mathbb{Z}}\Big[\theta\big(t-(nP+T_0)\big)-\theta\big(t-(nP+T_0)-T_1\big)\Big],\qquad P=T_0+T_1.
\end{align*}

In this section, we will now focus on a continuous drive protocol instead of a square wave drive. For this we will consider $f_1(t) = 1$ and,
\begin{align}
    f(t) = f_0\cos(\omega_D t) + \delta_f,
\end{align}
where $f_0, \omega_D$ denote the drive amplitude and frequency, respectively. Here $\delta_f$ is a constant shift parameter. For the continuous drive one can define the effective one cycle unitary drive operator $U_F$ as the evolution operator of one oscillation:
\begin{align}  \label{eq:Uf continuous}
    U_F \equiv U(T) =  \mathcal{T}\exp\!\left(-i\int_0^T H(t)\,dt\right), \qquad U_F^n = U(n T),
\end{align}
where $T$ is the time period of the drive oscillations: $T = 2 \pi/ \omega_D$.

In \cite{Das:2021gts}, analytical progress was made using the floquet perturbation theory (FPT). Till the first order in FPT one obtains,
\begin{align}
    U(nT) =  \begin{pmatrix}
a_n & b_n \\
c_n & d_n
\end{pmatrix}, ~
z_n = \frac{a_n z + b_n}{c_n z + d_n},
\end{align}
where, the matrix elements are:
\begin{align}
	a_n = \cos(n\theta) - i \frac{\sin(n\theta)}{\sqrt{1-\alpha^2}}, \,  b_n = i \frac{\alpha \sin(n\theta)}{\sqrt{1-\alpha^2}}, d_n = a_n^*, \, c_n = b_n^*.
\end{align}
with,
\begin{align}  \label{eq:continuous parameters}
	\theta = s\sqrt{1-\alpha^2}, \qquad s = \arccos\!\left(\cos\!\left(\frac{\pi \delta_f T}{L}\right)\right), \qquad \alpha = \sum_{n=-\infty}^{\infty} 
	\frac{J_n\!\left(\dfrac{2\pi f_0}{L\omega_D}\right) T}
	{n\pi + \dfrac{\pi \delta_f T}{L}} .
\end{align}
For this continuous drive, the heating phase is identified by $ |\text{Tr} U(T) | > 2$, while the non-heating phase arises for $ |\text{Tr} U(T) | < 2$, with the phase boundary at $ |\text{Tr} U(T) | =2$. Using the aforementioned approximation of $U$, these conditions translate into: the heating phase for $\alpha^2 >1$, the non-heating phase for $\alpha^2 < 1$, and the critical phase at $\alpha^2 = 1$. Under this $U$, the operator evolves as,
\begin{align}
	(U_F^\dagger)^n O(z, \zb) U_F^n = \left( \frac{\partial z_n}{\partial z} \right)^h \left( \frac{\partial \zb_n}{\partial \zb} \right)^{\hb} O(z_n, \zb_n) = \frac{1}{(c_n z + d_n)^{2h}} \frac{1}{(\bar{c}_n z + \bar{d}_n)^{2\hb}}  O(z_n, \zb_n).
\end{align}
With this, the autocorrelation function is then (for $\hb = h$):
\begin{align}  \label{eq:Gn continuous}
	G_n = (K_{0}|U_{F}^{n} | K_{0} ) = \frac{1}{|d_n|^{4 h}} =
	\begin{cases}
		\left(\dfrac{\sin ^2(n \theta )}{1-\alpha ^2}+\cos ^2(n \theta) \right)^{-2 h} ~~ \text{for } \alpha^2 < 1, \\
		\left( \dfrac{\sinh ^2\left(n \theta '\right)}{\alpha ^2-1}+\cosh ^2\left(n \theta '\right) \right)^{-2 h} ~~ \text{for } \alpha^2 > 1,
	\end{cases}
\end{align}
where, $\theta' = s \sqrt{\alpha^2 - 1}$. 

\begin{figure}[h]
	\centering
	\begin{subfigure}{0.45\textwidth}
		\centering
		\includegraphics[width=\linewidth]{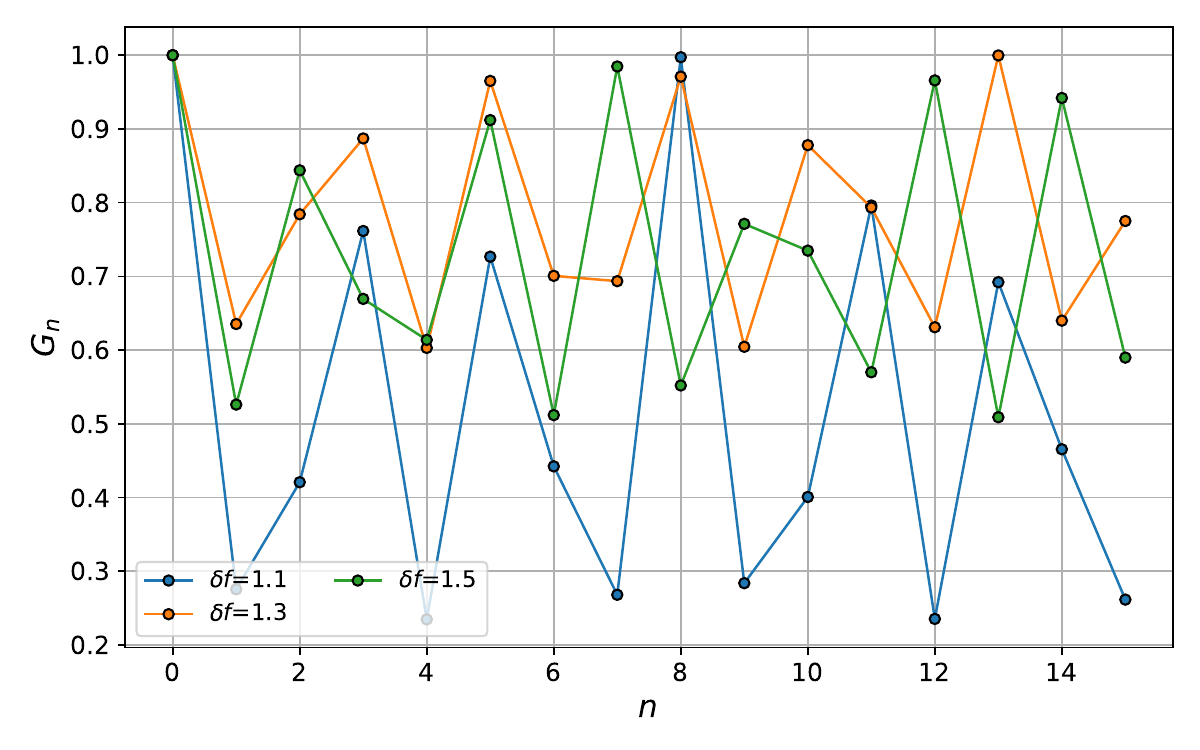}
		\caption{$\omega = 2, \, L = \pi, \, f_0 = 10, \, h=1$}
	\end{subfigure}
	\hfill
	\begin{subfigure}{0.45\textwidth}
		\centering
		\includegraphics[width=\linewidth]{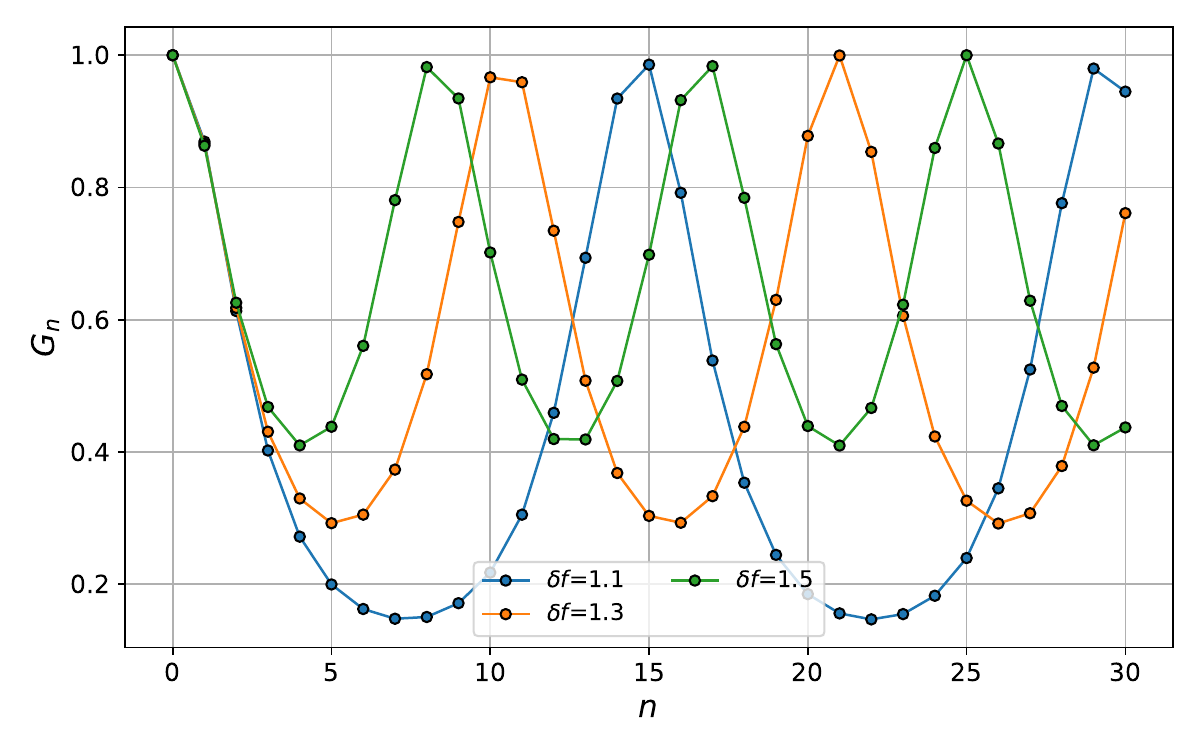}
		\caption{$\omega = 20, \, L = \pi, \, f_0 = 10, \, h=1$}
	\end{subfigure}
	\caption{$G_n$ as a function of $n$ in the nonheating phase of the continuous drive.}
	\label{fig:Gn nonheating continuous}
\end{figure}

Hence, in the nonheating phase $G_n$ shows oscillatory behaviour as a function of $n$ (see Fig.\ref{fig:Gn nonheating continuous}). The time period of this oscillation in terms of drive cycle is then: $n_{\text{period}} = \lceil \pi / \theta \rceil $. 

\begin{figure}[htbp]
	\centering
	\begin{subfigure}{0.45\textwidth}
		\centering
		\includegraphics[width=\linewidth]{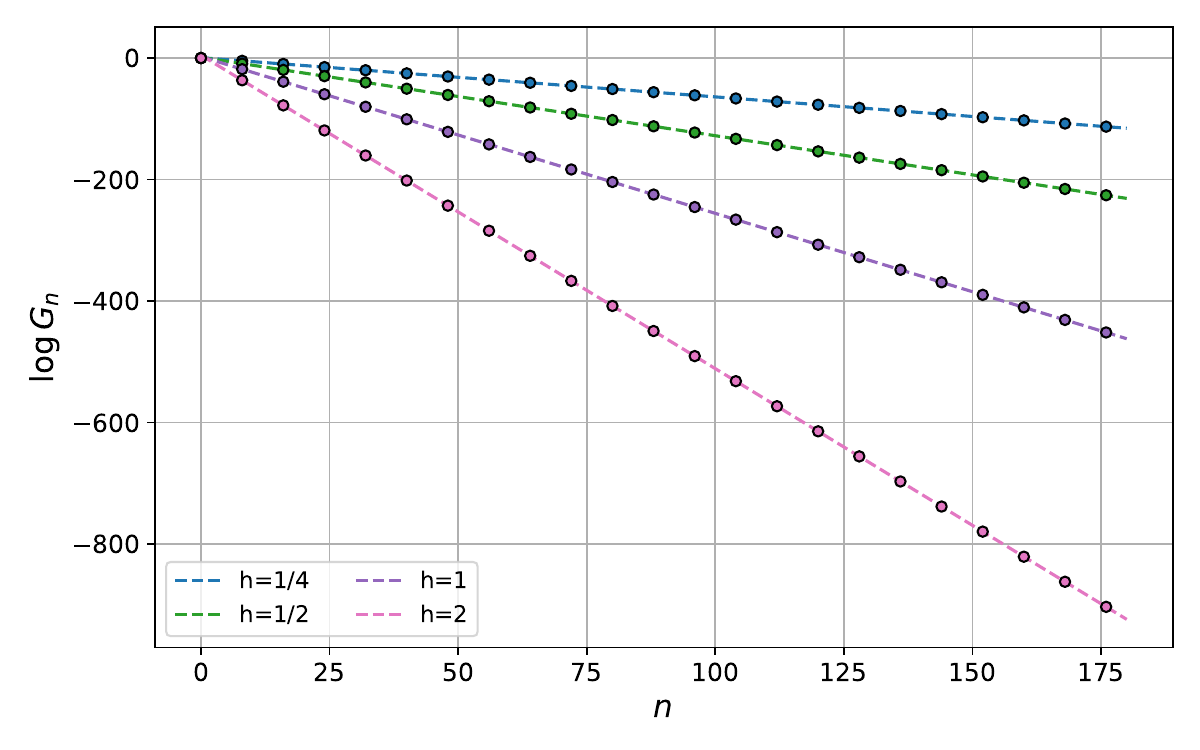}
		\caption{$\omega = 2, \, L = \pi, \, f_0 = 10, \, \delta_f=0.1$}
	\end{subfigure}
	\hfill
	\begin{subfigure}{0.45\textwidth}
		\centering
		\includegraphics[width=\linewidth]{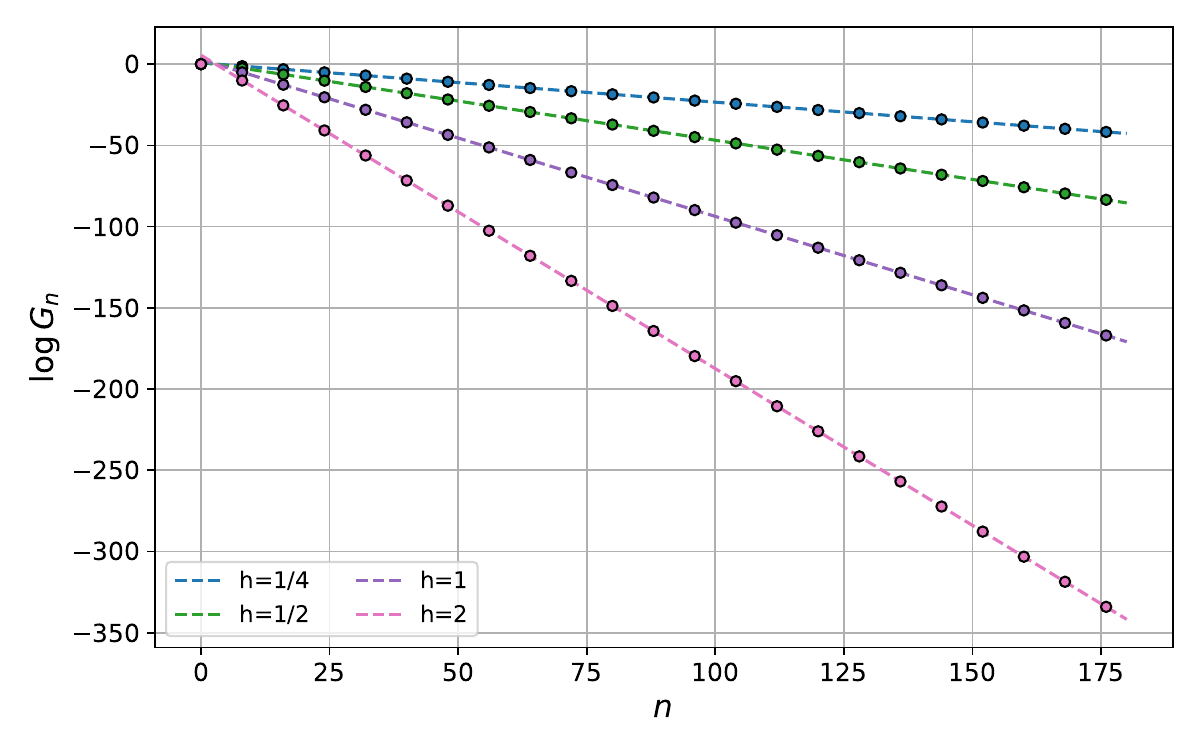}
		\caption{$\omega = 20, \, L = \pi, \, f_0 = 10, \, \delta_f=0.1$}
	\end{subfigure}
	\caption{$G_n$ as a function of $n$ in the heating phase of the continuous drive. The dots show the exact values, while the dashed lines show the approximation mentioned in \eqref{eq:Gn heating continuous approximation}.}
	\label{fig:Gn heating continuous}
\end{figure}

 In the heating phase $G_n$ decays with $n$ (see Fig.\ref{fig:Gn heating continuous}). Since $G_n$ is exponentially suppressed for large $\theta'$ or $n$ in the heating phase, one can then approximate:
\begin{align}  \label{eq:Gn heating continuous approximation}
	G_n \simeq e^{-4 h n \theta'} \, 4^h \left( \frac{\alpha^2}{\alpha^2 - 1} \right)^{-2h} .
\end{align}
Hence $\log G_n$ has a slope of $-4 h \theta'$ as a function of $n$. Fig.\ref{fig:Gn heating continuous} shows this approximation along with the exact behaviour of $G_n$.

\subsection{Arnoldi coefficients}

Similar to the suqare wave protocol, one can compute the Arnoldi coefficients $h_{n,n-1}$ following the algorithm described in $\S$\ref{sec:K-complexity}, with the unitary evolution operator described in \eqref{eq:Uf continuous}. 

\begin{figure}[hbtp]
	\centering
	\begin{subfigure}{0.45\textwidth}
		\centering
		\includegraphics[width=\linewidth]{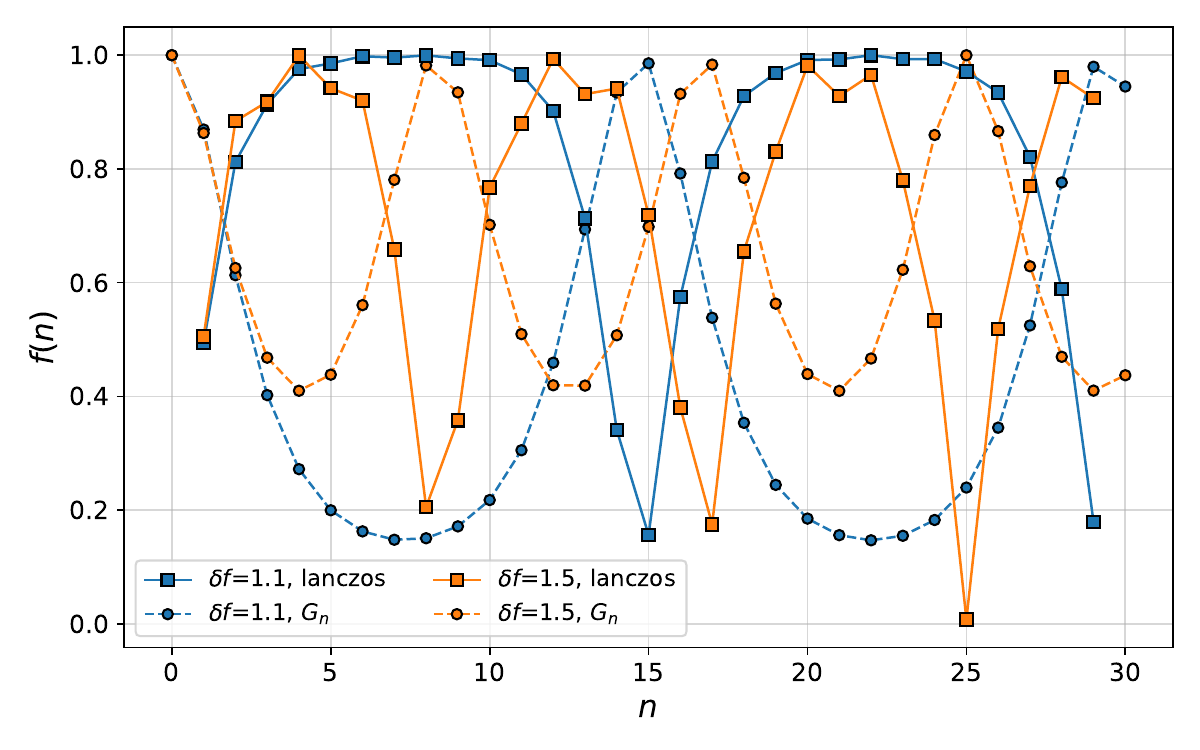}
		\caption{$\omega = 20, \, L = \pi, \, f_0 = 10, \, h=1$}
	\end{subfigure}
	\hfill
	\begin{subfigure}{0.45\textwidth}
		\centering
		\includegraphics[width=\linewidth]{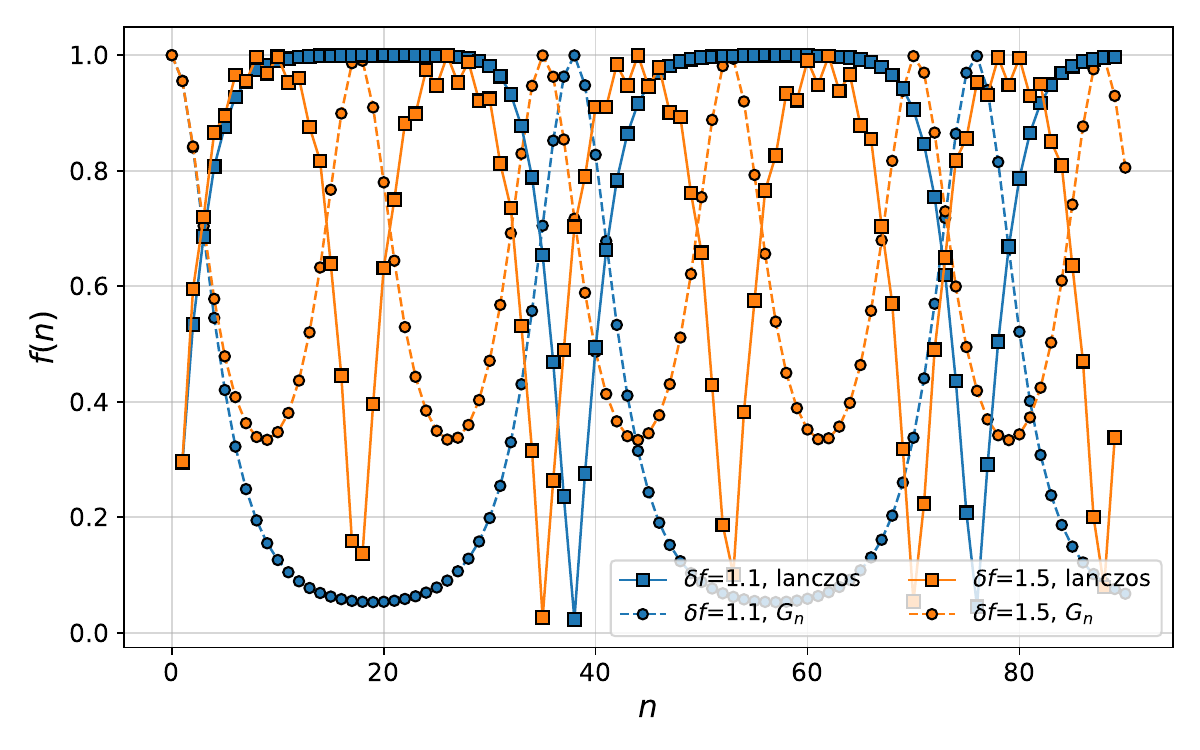}
		\caption{$\omega = 40, \, L = \pi, \, f_0 = 10, \, h=1$}
	\end{subfigure}
	\caption{Oscillations in $h_{n,n-1}$ as a function of $n$ in the non-heating phase of the continuous drive follow the frequency of $G_n$. The solid lines denote $h_{n,n-1}$, while the dashed lines of the same colour denote $G_n$ for the same parameters.}
	\label{fig:Lanczos nonheating continuous}
\end{figure}

Since the autocorrelation $G_n$ is oscillatory in the non heating phase, the resulting Arnoldi coefficients show oscillatory behaviour as a functiuon of drive cycle $n$. Interestingly, the frequency of oscillation of $h_{n,n-1}$ appears to follow the frequency of the $G_n$ oscillations, which has a drive cycle period of $n_{\text{period}} = \lceil \pi / \theta \rceil $. See Fig.\ref{fig:Lanczos nonheating continuous} for the corresponding examples.

In the heating phase, the Arnoldi coefficients fall exponentially with $n$, following the exponential decay of $G_n$. But unlike the square wave protocol, the expression of $G_n$ in \eqref{eq:Gn continuous} does not simplify to just an exponential decaying factor for arbitrary parameters $\omega, \delta_f$, i.e., $\theta'$ might not be large enough for the hyperbolic trigonometric factors to approximate an exponential form. Moreover, due to the complexity of the Lanczos Algorithm, the autocorrelations $G_m$ appear non-trivially in the expression of $h_{n,n-1}$; hence it becomes increasingly difficult to analytically express the Arnoldi coefficients in terms of $G_n$ for increasing $n$.

\begin{figure}[hbtp]
	\centering
	\begin{subfigure}{0.45\textwidth}
		\centering
		\includegraphics[width=\linewidth]{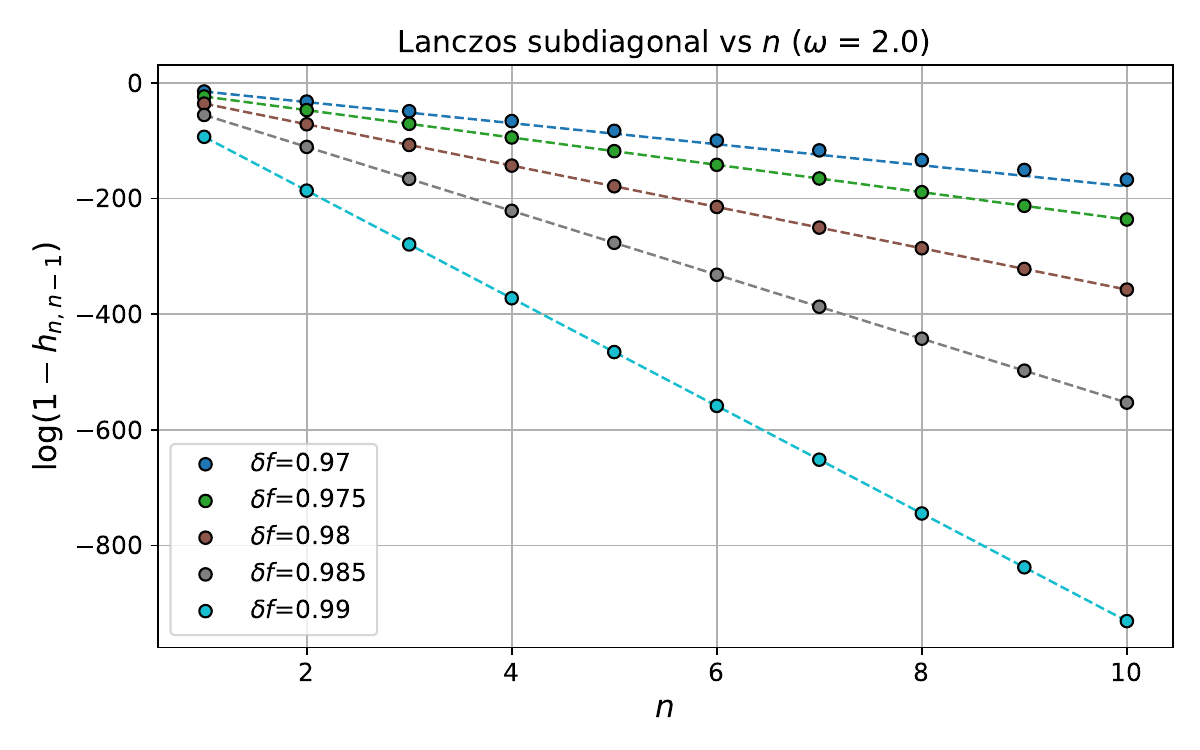}
		\caption{$\omega = 2, \, L = \pi, \, f_0 = 10, \, h=1$}
	\end{subfigure}
	\hfill
	\begin{subfigure}{0.45\textwidth}
		\centering
		\includegraphics[width=\linewidth]{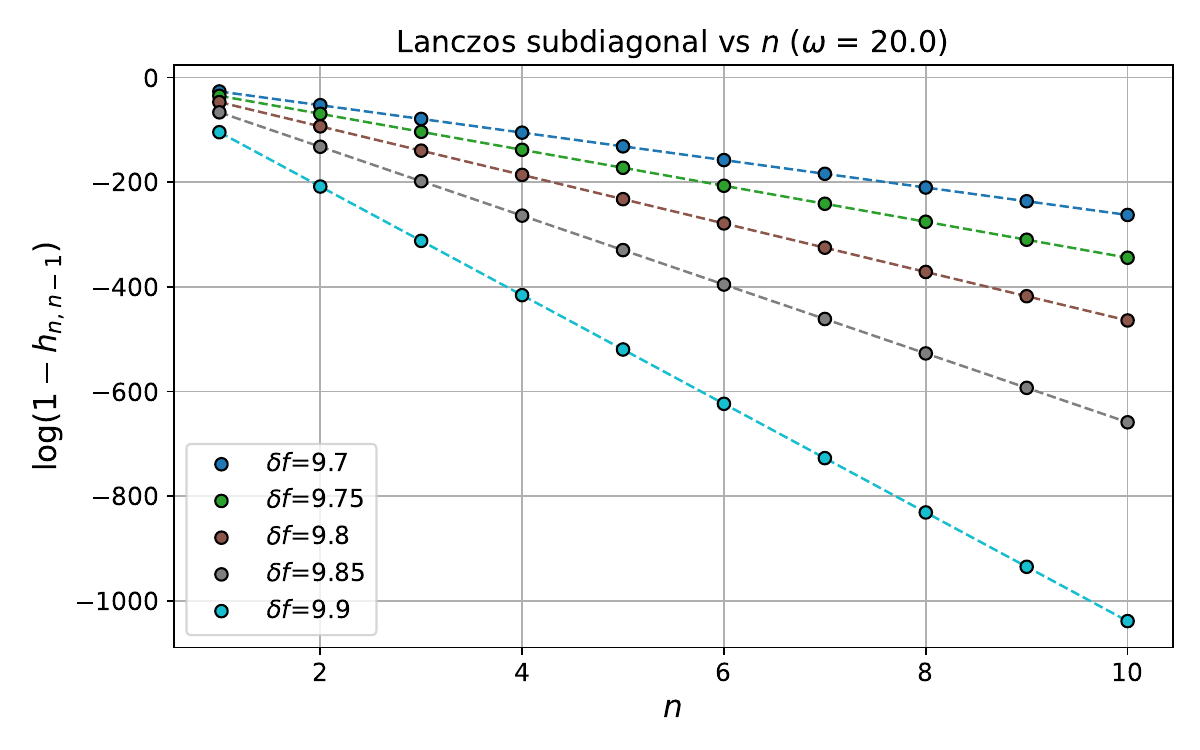}
		\caption{$\omega = 20, \, L = \pi, \, f_0 = 10, \, h=1$}
	\end{subfigure}
	\caption{$h_{n,n-1}$ as function of $n$ in the heating phase of continuous drive. The dots denote the exact coefficients, while the dashed lines denote the approximation shown in \eqref{eq:lanczos approx continuous heating}. The plots have been done for $n_0$ close to 1, i.e., $\delta_f \simeq \omega/2$ (see \eqref{eqn n0}). }
	\label{fig:Lanczos heating continuous}
\end{figure}

However, if $\theta' = s \sqrt{\alpha^2 - 1}$ is large enough, analytical approximations can be made following the exponential decaying form of $G_n$ shown in \eqref{eq:Gn heating continuous approximation}. This happens when the denominator in the expression of $\alpha$ in \eqref{eq:continuous parameters} becomes small, i.e., when the drive parameters $\delta_{f}, \omega$ follow the following relation:
\begin{align} \label{eqn n0}
n_0 \simeq \frac{\delta_f \, T}{L} = \frac{2 \pi \delta_f}{\omega L}, ~ n_0 \in \mathbb{Z}.
\end{align}
At the equality of the above relation $\alpha$ (i.e., $\theta'$) blows up. In this regime, the Arnoldi coefficients exhibit the following behaviour:
\begin{align} \label{eq:lanczos approx continuous heating}
	h_{n,n-1} \simeq 1 - \frac{1}{2} A^2 (A-1)^{2 (n-1)} e^{-8 h n \theta '}, ~ \text{where, } A = \left(4 \left(1-\frac{1}{\alpha^2}\right)\right)^{2 h} .
\end{align}
In Fig.\ref{fig:Lanczos heating continuous}, the Arnoldi coefficients have been plotted along with the above approximation for $n_0 \simeq 1$.

\section{Continuous drive in critical fermions}
\subsection{Floquet Krylov dynamics with open boundary conditions}

We now discuss the lattice realization of the continuously driven CFT introduced in the previous section, considering a system with open boundary conditions. The corresponding time-dependent Hamiltonian takes the form~\cite{Katsura:2011ss,Katsura:2011zyx,PhysRevB.84.165132,Das:2021gts}
\begin{equation}
H(t)= J(t) H^{\mathrm{lat}}_0 + J_1 H_q ,
\end{equation}
where the uniform hopping term \(H^{\mathrm{lat}}_0\) is defined in Eq.~\eqref{eq:Hlat}, while
\begin{equation}
H_q = \sum_{i=1}^{L-1} 
\cos\!\left(\frac{2\pi(i+\tfrac{3}{2})}{L+1}\right)
\left(c_i^\dagger c_{i+1} + \mathrm{h.c.}\right)
\end{equation}
introduces a spatial modulation of the hopping amplitudes. The drive enters through the time-dependent coupling
\begin{equation}
J(t)=\delta_f + J_0 \cos(\omega_D t),
\end{equation}
with driving frequency \(\omega_D\). The Floquet operator corresponding to one driving period \(T=2\pi/\omega_D\) is obtained by time-ordered evolution over the period,
\begin{equation}
U_F = \mathcal{T}\exp\!\left(-i\int_0^T H(t)\,dt\right).
\end{equation}
In the numerical implementation the time evolution is approximated using a midpoint Trotter discretization of the Floquet period. To ensure numerical stability of the moment sequence and the subsequent Lanczos procedure, all computations are carried out using arbitrary-precision arithmetic implemented with the \texttt{python-flint} library~\cite{flint,flint2}.

As in the previous subsection, the system is initialized in the half-filled Fermi sea of the static Hamiltonian \(H^{\mathrm{lat}}_0\). For open boundary conditions the single-particle eigenmodes are given analytically by
\begin{equation}
\phi_k(j) = 
\sqrt{\frac{2}{L+1}}\,
\sin\!\left(\frac{\pi k j}{L+1}\right),
\qquad
k=1,\dots,L,
\end{equation}
and the initial many-body ground state is obtained by occupying the lowest \(L/2\) modes. The return amplitude
\begin{equation}
G_m = \langle G | U_F^m | G \rangle
\end{equation}
is evaluated using the determinant formula introduced previously (\ref{eq:return_lat}).

To analyze the Krylov growth generated by the Floquet dynamics we construct the Krylov basis from the sequence of moments. The Lanczos Algorithm applied to these moments yields the Arnoldi coefficients \(h_{n,n-1}\), which encode the spreading of the state in Krylov space.

\begin{figure}[t]
\centering
\begin{subfigure}{0.45\textwidth}
\centering
\includegraphics[width=\linewidth]{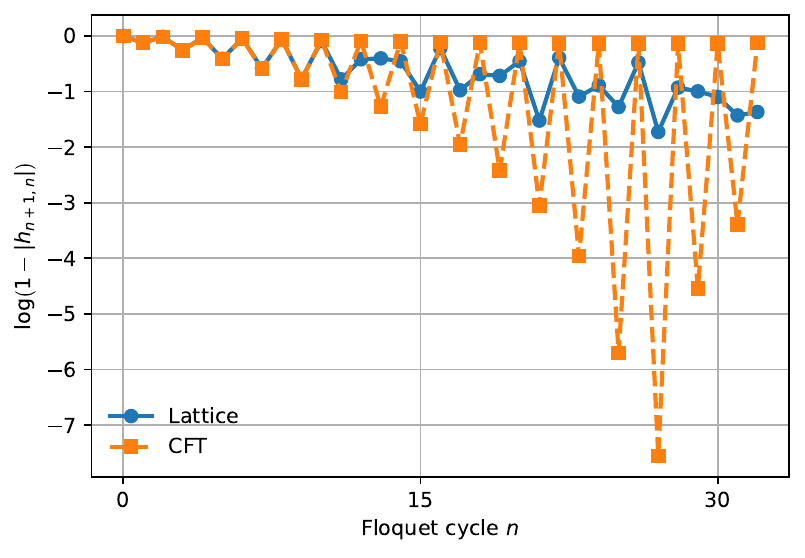}
\caption{Non-heating phase ($\delta_{f} , \omega_{D}) = (15,40)$.}
\end{subfigure}
\hfill
\begin{subfigure}{0.45\textwidth}
\centering
\includegraphics[width=\linewidth]{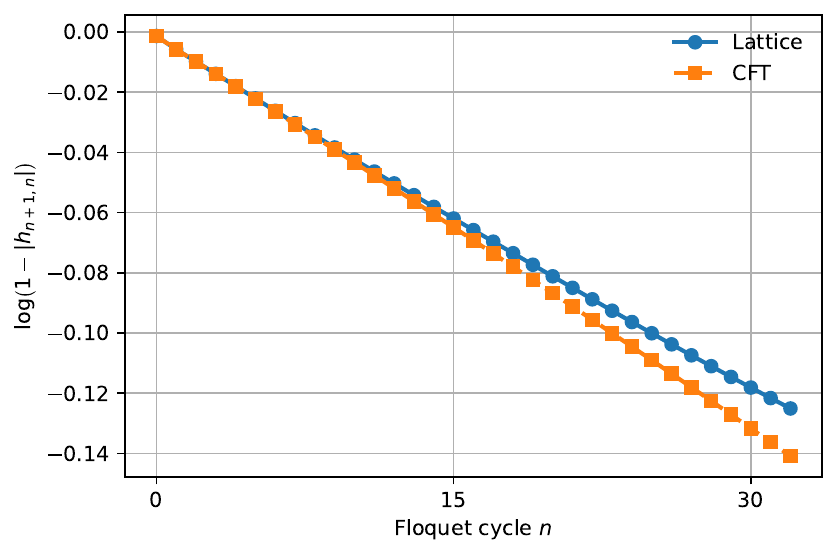}
\caption{Heating phase ($\delta_{f} , \omega_{D}) = (0.1,40)$}
\end{subfigure}
\caption{Behaviour of $h_{n,n-1}$ in different dynamical regimes of the Floquet drive.}
\label{fig:contobc_krylov}
\end{figure}

Figure \ref{fig:contobc_krylov} show the behaviour of the Arnoldi coefficients \(h_{n,n-1}\) obtained from the continuous driving protocol. It shows improved agreement between the lattice and CFT predictions over a significantly larger number of Floquet cycles compared to the square-wave drive considered previously. In the non-heating phase the coefficients exhibit an oscillatory behaviour as a function of the Krylov index, reflecting the periodic dynamics of the return amplitude. In contrast, in the heating phase the coefficients approach unity exponentially, and eventually the decay convergence slows down in the lattice calculation, indicating rapid spreading in Krylov space. The system size used in the numerics, here \(L=128\). 

We also study the same Floquet dynamics with periodic boundary conditions. In this case the single–particle eigenmodes of the uniform Hamiltonian are plane waves,
\begin{equation}
\phi_k(j)=\frac{1}{\sqrt{L}}\,e^{2\pi i k j/L}, 
\qquad k=0,\dots,L-1,
\end{equation}
and the half–filled Fermi sea is obtained by occupying the lowest energy momentum modes. In order to match the CFT results in this case also we consider an excited initial state obtained by adding two lowest–lying excitations above the Fermi sea~\cite{Berganza:2011mh}. The Floquet operator is constructed in the same manner as in the open boundary case.

\begin{figure}[t]
\centering
\begin{subfigure}{0.45\textwidth}
\centering
\includegraphics[width=\linewidth]{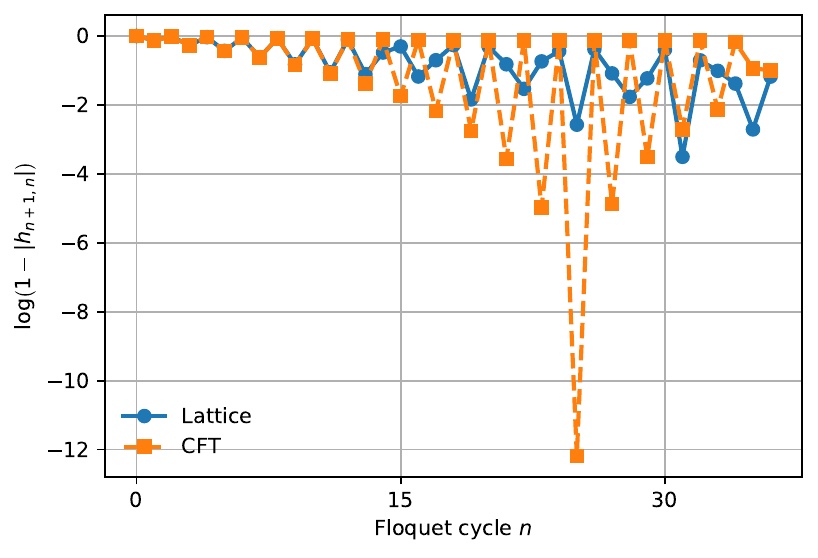}
\caption{Non-heating phase ($\delta_{f} , \omega_{D}) = (15,40)$.}
\end{subfigure}
\hfill
\begin{subfigure}{0.45\textwidth}
\centering
\includegraphics[width=\linewidth]{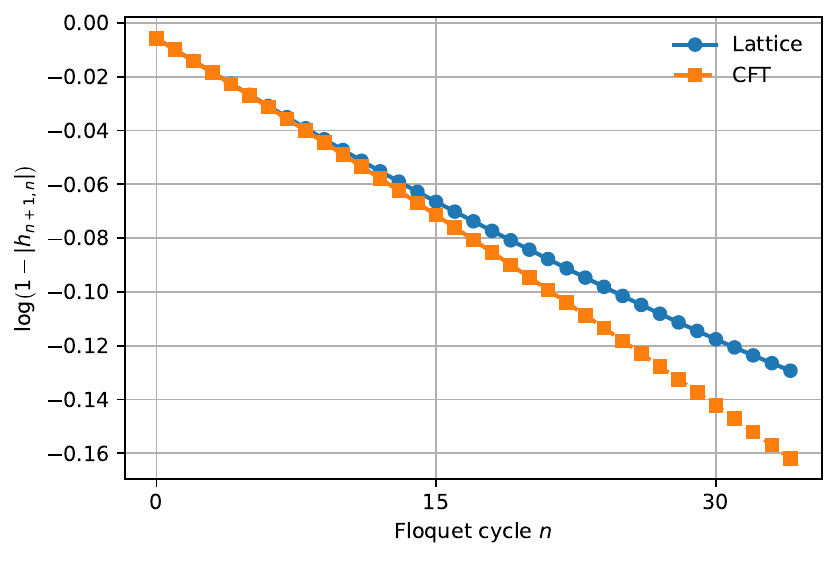}
\caption{Heating phase ($\delta_{f} , \omega_{D}) = (0.1,40)$}
\end{subfigure}
\caption{Behaviour of $h_{n,n-1}$ in different dynamical regimes of the Floquet drive with Periodic Boundary Conditions.}
\label{fig:cont_krylov}
\end{figure}

The Krylov construction Figure \ref{fig:cont_krylov} then proceeds from the sequence of return amplitudes \(G_m\) computed from the determinant overlap formula. Although the initial state differs from the open boundary case, the qualitative behaviour of the Arnoldi coefficients \(h_{n,n-1}\) remains unchanged. In particular, the non-heating regime exhibits oscillatory behaviour of the coefficients, while in the heating regime they rapidly approach unity.

\subsection{K-complexity of Correlation Matrix}
We now consider the K-complexity of Correlation Matrix for the continuous driving protocol. For each choice of driving strength $\delta_f$, we compute the averaged sub-diagonal coefficients $h^{C}_{n,n-1}$ and the Krylov complexity $C_K(j)$ by averaging over $n_{\mathrm{samples}}=10$ realizations with $\sim 5\%$ uniform fluctuations in $\delta_f$ around its base value and with fixed $\omega_D = 40$.

\begin{figure}[t]
\centering
\includegraphics[width=0.48\linewidth]{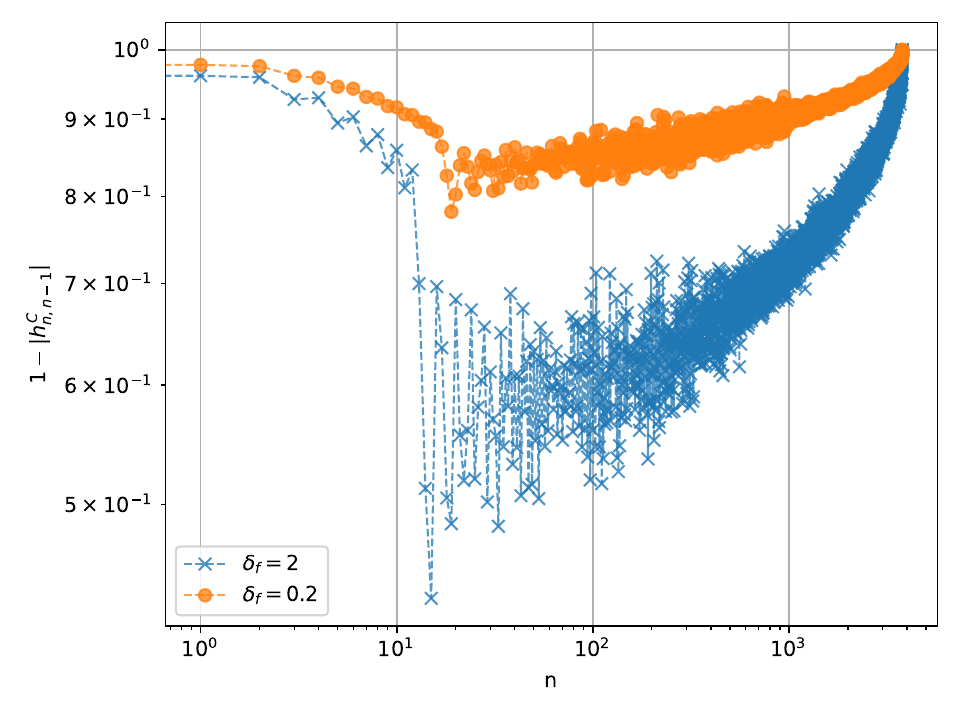}
\hfill
\includegraphics[width=0.48\linewidth]{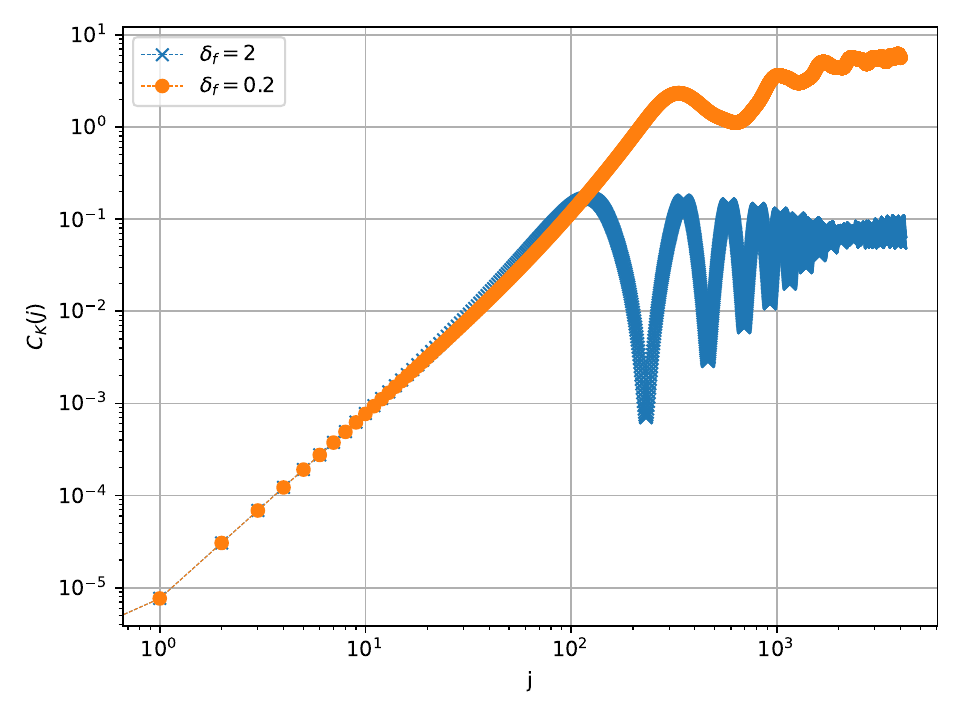}
\caption{
Left: $1 - |h^{C}_{n,n-1}|$ vs $n$ (log--log). 
Right: Krylov complexity $C_K(j)$ vs $j$ (log--log). 
Results are shown for $\delta_f = 2$ and $\delta_f = 0.2$.
}
\label{fig:full_cont_krylov}
\end{figure}

Figure~\ref{fig:full_cont_krylov} shows the behavior for two representative values, $\delta_f = 2$ and $\delta_f = 0.2$. In the both the heating ($\delta_f = 0.2$) and non-heating phase ($\delta_f = 2$), the sub-diagonal growth at small $n$ and the growth is considerably weaker compared to the square-wave drive case. This suppressed growth is followed by oscillations around nearly a constant value, indicating the formation of a pronounced Lanczos plateau. Interestingly, in the non-heating phase the plateau forms at slightly higher values than in the heating phase. At later Krylov steps, the coefficients decay toward zero, marking the onset of the Lanczos descent due to finite system size.

The K-complexity, unlike in the square-wave drive case, behaves nearly identically at early Floquet steps for both the phases. At later times, it develops oscillations in the non-heating phase, whereas in the heating phase the growth continues with only minor oscillations.

\subsection{Spectral Statistics}

To further characterize the dynamical regimes of the driven lattice model we analyze the spectral statistics of the Floquet operator. The eigenvalues of the Floquet operator \(U_F\) can be written as
\begin{equation}
U_F |\psi_\alpha\rangle = e^{-i\theta_\alpha} |\psi_\alpha\rangle ,
\end{equation}
where \(\theta_\alpha\) are the quasienergies defined modulo \(2\pi\). 

The statistical properties of the spectrum are quantified using the adjacent level–spacing ratio with $\varepsilon_\alpha = -\theta_\alpha$
\begin{equation}
r_\alpha = 
\frac{\min(s_\alpha,s_{\alpha+1})}{\max(s_\alpha,s_{\alpha+1})},
\qquad
s_\alpha = \varepsilon_{\alpha+1}-\varepsilon_\alpha .
\end{equation}

In the numerical calculations the Floquet operator is constructed by discretizing the time evolution over one driving period using a midpoint Trotter scheme. At each time step the instantaneous Hamiltonian
\begin{equation}
H(t)= J(t)H_0 + J_1 H_q ,
\qquad
J(t)=\delta_f + J_0 \cos(\omega_D t)
\end{equation}
is diagonalized and exponentiated. The full Floquet operator is obtained as the product of the short–time propagators over the period.

We study the level statistics as a function of the driving parameter \(\delta_f\) while keeping the driving frequency fixed at \(\omega_D=40\). The analysis is performed for several system sizes \(L=122,\,242,\) and \(482\). For each value of \(\delta_f\) the quasienergy spectrum is obtained from the eigenphases of \(U_F\), and the corresponding \(r\)-factor is computed.

\begin{figure}[t]
\centering
\includegraphics[width=0.45\textwidth]{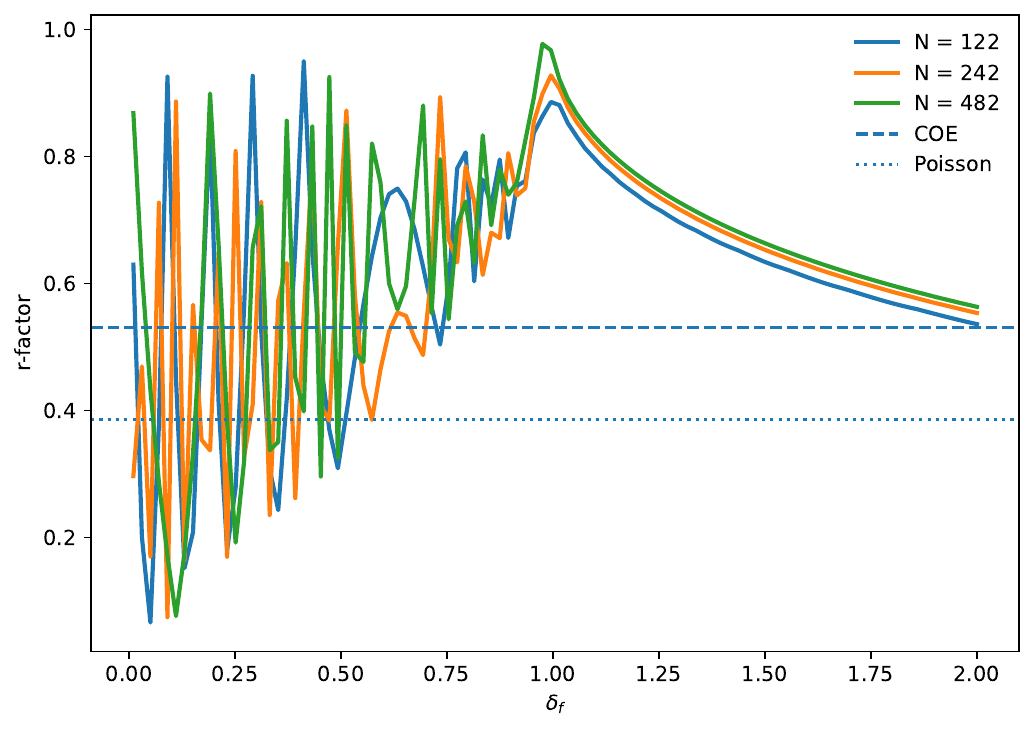}
\caption{Average adjacent level-spacing ratio \(\langle r\rangle\) of the effective Floquet Unitary operator as a function of the driving parameter \(\delta_f\) at fixed frequency \(\omega_D =40\). The results are shown for several system sizes. The rapid oscillations of \(\langle r\rangle\) in the heating regime and its smooth variation in the non-heating regime indicate a change in the spectral properties of the Floquet spectrum across the transition.}
\label{fig:r_heff_scan}
\end{figure}

The resulting behaviour of \(\langle r\rangle\), shown in Fig.~\ref{fig:r_heff_scan}, reveals a clear change in the spectral properties as the driving parameter \(\delta_f\) is varied. In the non-heating regime (\(\delta_f>1\)), the \(r\)-factor varies smoothly as \(\delta_f\) increases. In contrast, in the heating regime (\(\delta_f<1\)) the \(r\)-factor exhibits rapid oscillations as a function of \(\delta_f\). In both regimes the r -factor does not approach the universal limits associated with either Poisson or Wigner--Dyson statistics, indicating that the Floquet spectrum of the effective Hamiltonian remains non-generic in both phases. Moreover, as the system size \(N\) increases, the value of \(\langle r\rangle\) at the phase transition points tends towards unity, reflecting the increasing rigidity of the quasienergy spectrum in the large-system limit.

\subsection{Transition Graph Representation}

\begin{figure}[!htbp]
\centering
\includegraphics[width=0.45\textwidth]{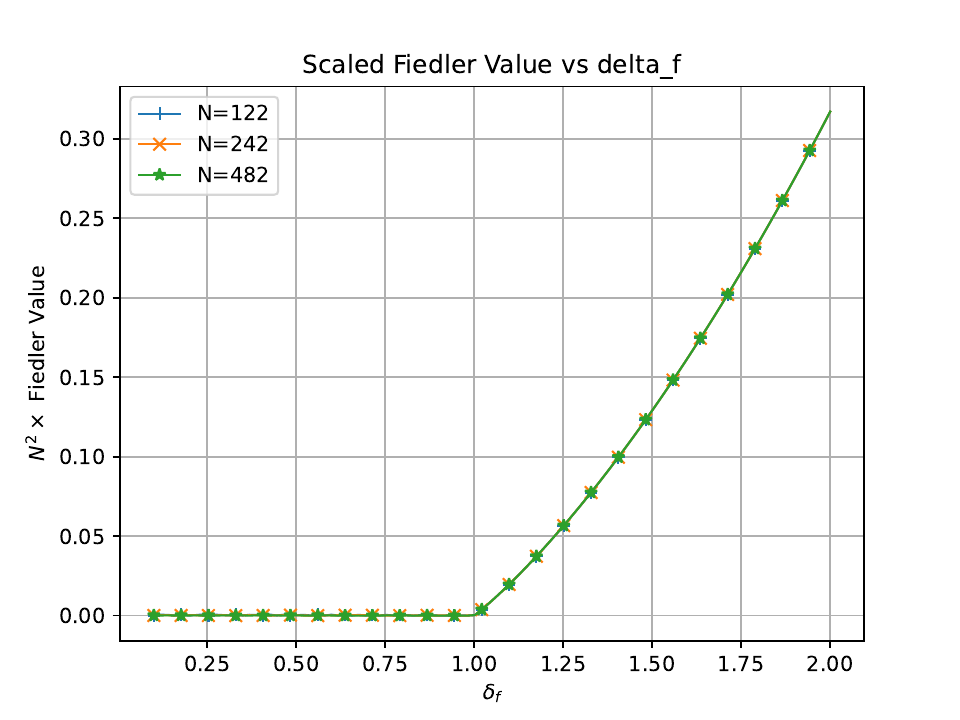}
\caption{Fiedler value of the Laplacian $L_W$, rescaled by $N^2$, plotted as a function of the drive parameter $\delta_f$ for different system sizes.}
\label{fig:filder_scan_U}
\end{figure}

From the unitary \(U_F\), we construct a stochastic transition matrix~\cite{2001JPhA...34L.319B,2001JPhA...34.8485T,Bressanini:2022qrr}
\begin{equation}
W_{ij} = \left|(U_F)_{ij}\right|^2,
\end{equation}
which defines a weighted graph over the lattice sites. By construction, \(W\) is doubly stochastic, since
\begin{equation}
\sum_i W_{ij} = 1,
\qquad
\sum_j W_{ij} = 1.
\end{equation}
In our numerics, we find that \(W\) is symmetric up to numerical precision, \(W_{ij} \approx W_{ji}\), with deviations of order \(10^{-13}\), reflecting an effectively undirected structure of the induced graph. To remove residual numerical artifacts arising from finite precision and time discretization, we explicitly symmetrize the matrix as
\begin{equation}
A = \frac{1}{2}\left(W + W^T\right),
\end{equation}
which we then use as the adjacency matrix for subsequent graph-theoretic analysis~\cite{Fiedler1973AlgebraicCO,chung1997spectral}. The corresponding graph Laplacian is
\begin{equation}
L_{W} = D - A,
\qquad
D_{ii} = \sum_j A_{ij}.
\end{equation}
Here, \(D\) is the degree matrix, with diagonal entries \(D_{ii}\) given by the weighted degree of site \(i\). This allows us to interpret the dynamics as a classical transition process on an undirected weighted graph. We compute the second-smallest eigenvalue of the graph Laplacian \(L_{W}\), known as the Fiedler value~\cite{Fiedler1973AlgebraicCO}, and plot it in Figure ~\ref{fig:filder_scan_U} as a function of the drive parameter \(\delta_f\). This provides a direct probe of how the connectivity of the Floquet-induced graph evolves with the driving strength.

\begin{figure}[t]
\centering

\includegraphics[width=0.45\textwidth]{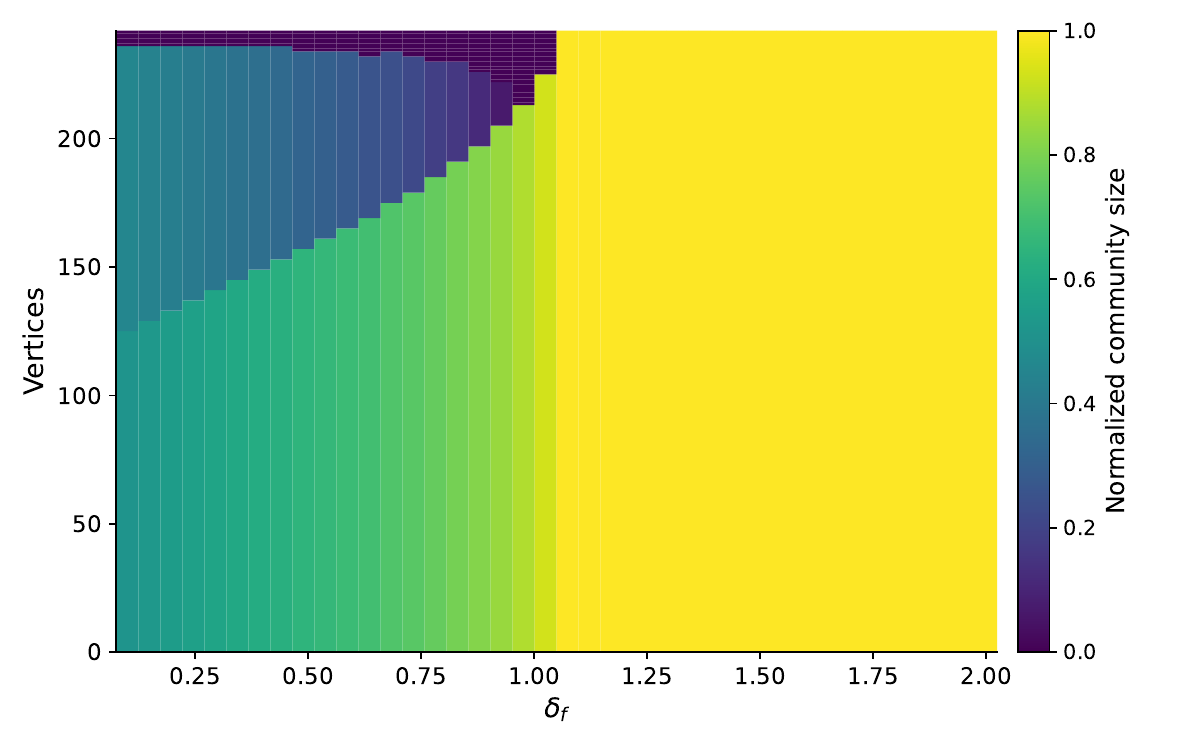}

\includegraphics[width=0.42\textwidth]{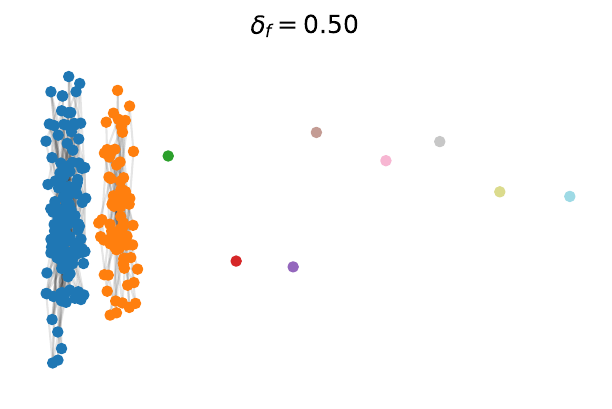}
\hfill
\includegraphics[width=0.42\textwidth]{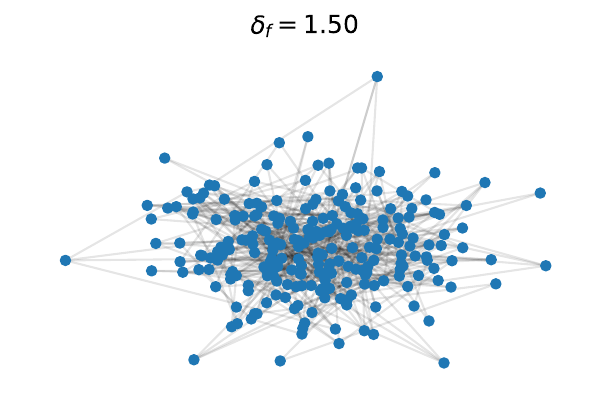}

\caption{Visualization of the Floquet-induced transition graph from $W_{ij}=|U_{ij}|^2$ together with its evolution across $\delta_f$. For $\delta_f<1$ (bottom-left), the network effectively fragments into two major clusters, while for $\delta_f>1$ (bottom-right) it forms a single connected cluster. The partition plot (top) shows the network structure with $\delta_f$, highlighting the transition from weakly separated to globally connected behavior.}
\label{fig:community_graphs}
\end{figure}

To further visualize the structure of the Floquet-induced transition graph, we construct an undirected weighted network from the symmetrized transition matrix $A$. We then threshold the network by removing edges with weight smaller than \(1/N^2\), and identify the connected components of the resulting graph. In
Figure~\ref{fig:community_graphs}, the resulting approximate groups of vertices are displayed as spatially separated clusters, where nodes belonging to the same group are grouped into visually distinct networks. Since the graph is constructed by thresholding weak transition probabilities, these groups should be interpreted as effective as they are weakly connected to one another rather than strictly isolated structures.

For $\delta_f > 1$ (i.e., in the non-heating phase), the graph is dominated by a single large cluster, reflecting global connectivity. In contrast, in the heating phase for $\delta_f < 1$, the network splits into two dominant clusters accompanied by several weakly connected single-vertex components. These visualizations provide an intuitive real-space representation of the connectivity properties inferred from the spectral diagnostics.

\begin{figure}
\centering

\includegraphics[width=0.5\textwidth]{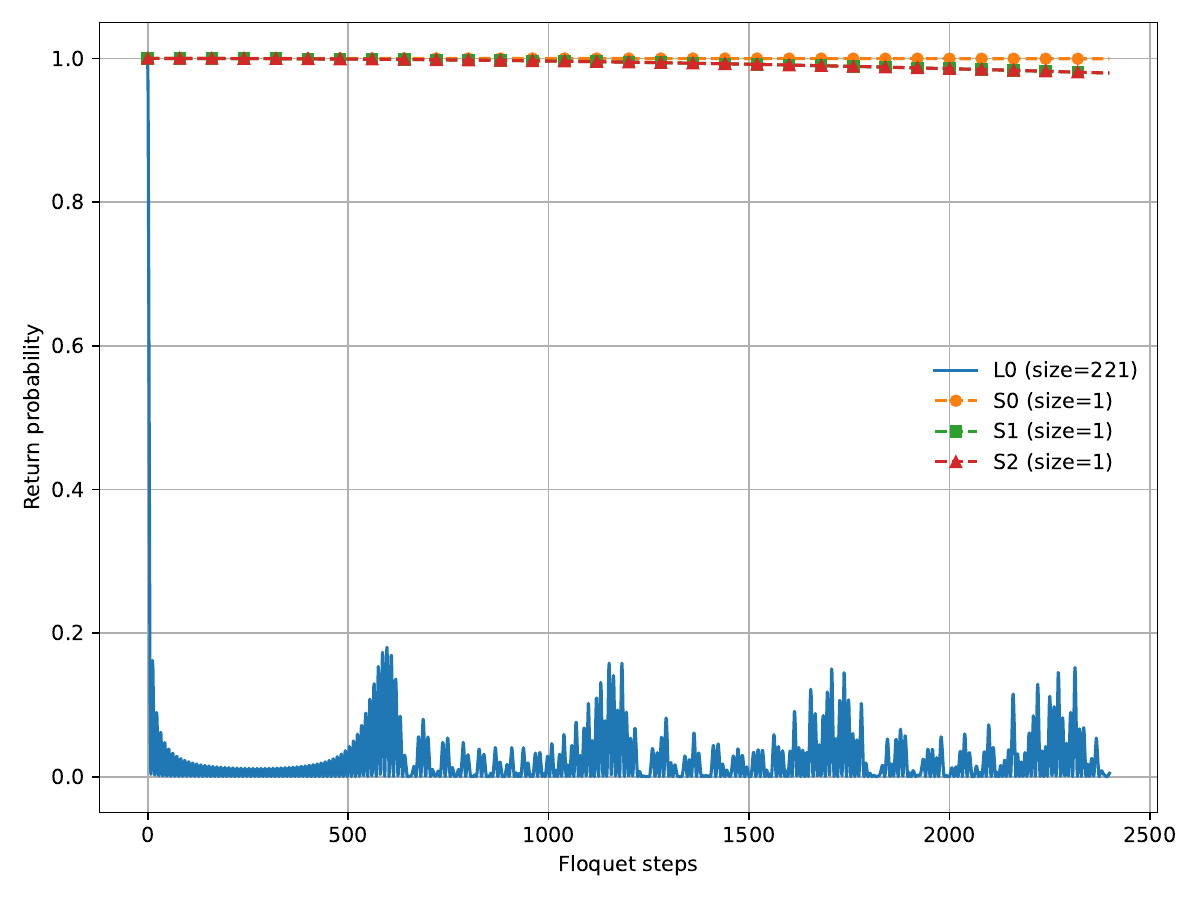}

\vspace{0.5cm}

\includegraphics[width=0.45\textwidth]{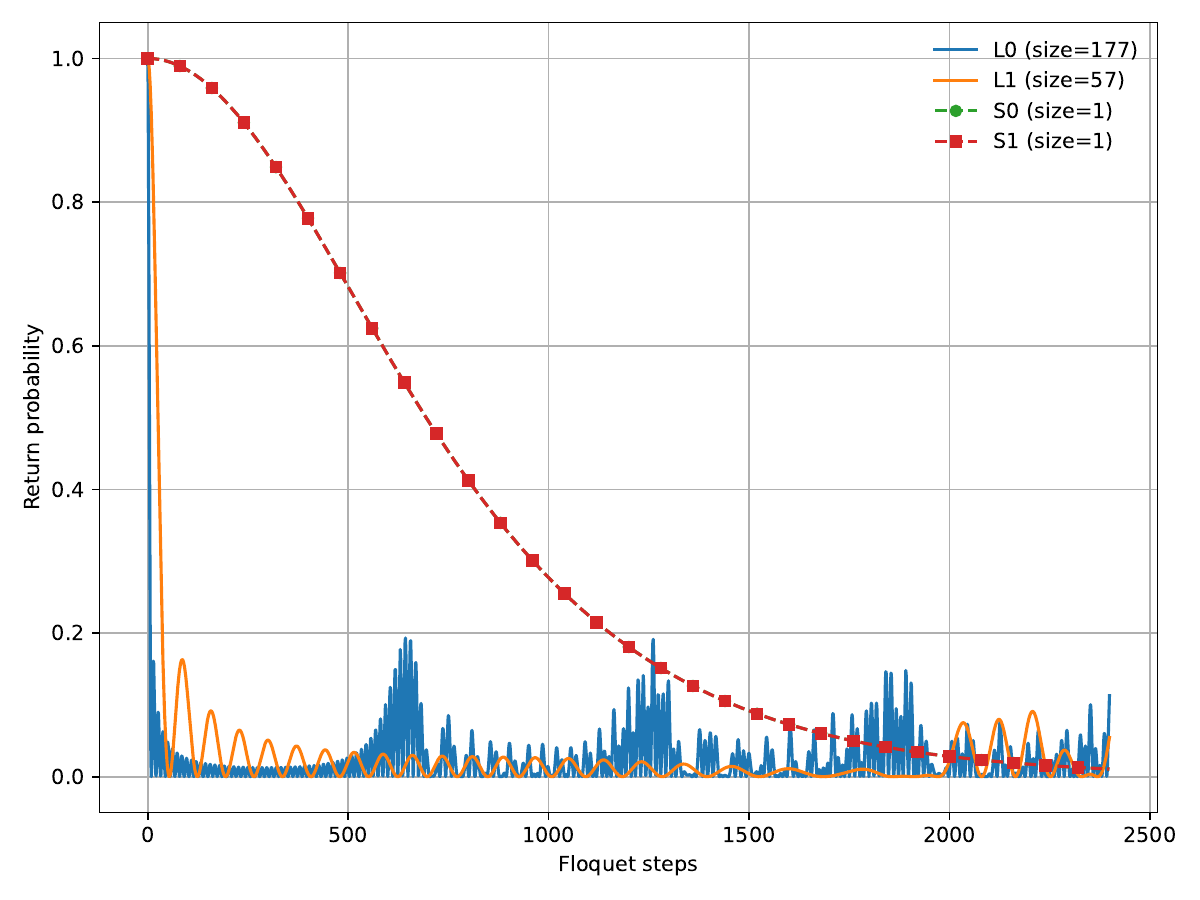}
\hfill
\includegraphics[width=0.45\textwidth]{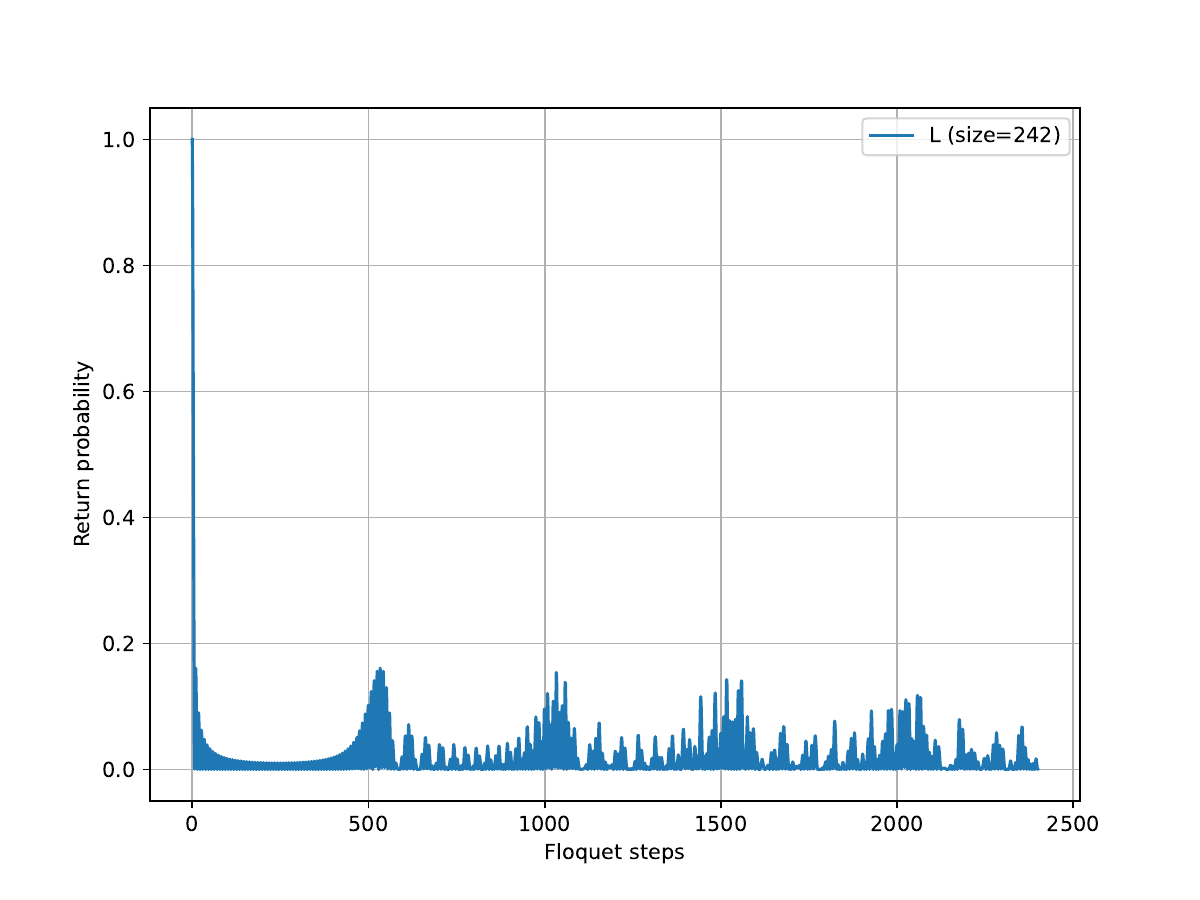}

\vspace{0.3cm}

(a) $\delta_f=0.7$ \hspace{0.25\textwidth}
(b) $\delta_f=1$ \hspace{0.25\textwidth}
(c) $\delta_f=1.5$

\caption{Return probability dynamics for various values of the drive parameter $\delta_f$. For $\delta_f<1$ (bottom left), slow-decaying modes associated with isolated sites are visible, indicating transport bottlenecks. Near $\delta_f = 1$ (top), the slowest mode exhibits a pronounced slowing down. For $\delta_f > 1$ (bottom right), there is only a single cluster.}
\label{fig:return_dynamics}
\end{figure}

To probe the dynamical implications of the graph structure, we plot the single-particle return probability (see Figure~\ref{fig:return_dynamics}):
\begin{equation}
P_i(t) = |\langle i | U_F^t | i \rangle|^2,
\end{equation}
starting from a localized state. Here $|i\rangle$ denotes a localized basis state with the particle initially occupying vertex $i$. For $\delta_f > 1$, $P_i(t)$ decays rapidly for all sites, indicating efficient spreading and global connectivity. In contrast, for $\delta_f < 1$, the dynamics becomes heterogeneous: sites in large communities decay quickly, while isolated sites exhibit slow decay. Near $\delta_f \sim 1$, the dynamics of isolated single-site modes exhibits the slowest relaxation.

We also find that, for the square-wave drive, the matrix \( W \) is no longer symmetric, \( W_{ij} \neq W_{ji} \), indicating that the effective graph becomes directed. Moreover, even the symmetrized component,
\begin{equation}
W^{(\mathrm{sym})}_{ij} = \tfrac{1}{2}(W_{ij} + W_{ji}),
\end{equation}
is largely insensitive to the transition between the heating and non-heating regimes. This suggests that the onset of heating is not captured by the connectivity of \( W \). We assume this is because in square-wave drives the drive parameters changes abruptly in time, which leads to stronger mixing between modes. In smooth drives they change gradually, allowing the system to retain more local structure.

\section{Discussions}

In this work, we have studied the notion of K-complexity in periodically driven conformal field theories. In particular, we study two different driving protocols: the square wave drive (characterized by two time parameters $T_0, T_1$) and the continuous sinusoidal drive (expressed in terms of the drive frequency $\omega_D$ and the drive strength $\delta_f$). Since we are interested in the evolution of the system at stroboscopic times, the corresponding Krylov basis is made out of the evolution operator of an entire time period ($U_F = e^{i H_F T}$), rather than the infinitesimal evolution of the operator through the effective Hamiltonian ($H_F$). In this basis the unitary evolution operator $U_F$ assumes an upper Hessenberg form, and we denote the sub-diagonal elements $h_{n,n-1}$. For chaotic dynamics, the sub-diagonal elements saturate to 1, and all the other elements of the matrix vanish. On the other hand, for integrable dynamics no such saturation is observed, and the sub-diagonal elements show oscillatory behaviour.

For both of the aforementioned driving protocols, two different phases are known: the heating phase and the non-heating phase. For the discrete square wave drive protocol the exact operator dynamics has been analytically calculated in \cite{Wen:2018agb}, while for the continuous drive analytical approximations of the operator dynamics were obtained using the floquet perturbation theory in \cite{Das:2021gts}. Using these analysis, we have been able to obtain the behaviour of the Arnoldi coefficients $h_{n,n-1}$ as a function of the drive cycle $n$ in both of the phases. For both driving protocols, in the heating phase, the Arnoldi coefficients saturate exponentially to 1 as a function of the drive cycles, where the slope of $\log (1 - h_{n,n-1})$ is proportional to both the scaling dimension $h$ and the drive cycle $n$. On the other hand, the Arnoldi coefficients exhibit oscillatory behaviour in the non-heating phase, whose frequency of oscillations follows the frequency of the autocorrelation function $G_n$. Although it is cumbersome to obtain the analytic expressions of $h_{n,n-1}$ for large $n$, we have been able to approximate the behaviour of the Arnoldi coefficients in the heating phase, which show excellent match with the exact coefficients.

We have also simulated the corresponding CFT dynamics on a lattice too, for both the discrete and continuous drives. Again in both cases, the autocorrelation $G_n$ and the Arnoldi coefficients $(1 - h_{n,n-1})$ decay exponentially in the heating phase, and in the non-heating phase they show oscillatory behaviour. This is similar to what one expects from the exact CFT dynamics. The lattice dynamics matches the CFT dynamics till the time period when the finite size effects start to show up.  In the lattice case we also study the K complexity of the correlation matrix. We find that the subdiagonal elements $h_{n,n-1}$ exhibit behavior similar to that of the Lanczos coefficients in finite spin-chain systems governed by time-independent Hamiltonians. However, the corresponding behavior of the Krylov complexity is different for the two drives.

Although the Arnoldi coefficients for the lattice behave similarly for both protocols, the underlying spectral dynamics show drastically different signatures. For a unitary drive, one can define an effective level spacing ratio $\vev{r}$ from the quasienergies $\varepsilon_{\alpha}$ obtained from the eigenvalues $e^{-i \varepsilon_\alpha}$ of $U_F$. In the discrete drive, we find $\vev{r}$ fluctuates about the Wigner-Dyson value of 0.53 in the heating phase, while in the non-heating phase it fluctuates about the Poisson value of 0.386. Thus, in the heating phase the lattice quasienergies exhibit random matrix characteristics, implying that the lattice dynamics truly becomes chaotic and the quasienergies are correlated through level repulsion. Similarly in the non-heating phase the lattice dynamics become closer to integrable case, resulting into approximately uncorrelated Poisson-like quasienergy distribution. This is drastically different from what we observe in the continuous drive, for which we observe that in the heating phase $\vev{r}$ shows erratic fluctuations; while in the non-heating phase $\vev{r}$ varies smoothly, but does not saturate to either of the Wigner-Dyson or Poisson values. This behaviour implies that in neither of the phases, the effective lattice dynamics can be characterized by the random matrix like features.

This qualitative difference of the two lattice drive protocols is reflected in the graph connectivity of the effective unitary transition matrix. In the continuous drive, \(U_F\) gives rise to a near-symmetric transition matrix that can be interpreted as an undirected graph, whose connectivity reorganizes across the transition as reflected in the Fiedler value and the structure of its connected components after thresholding. In contrast, the square-wave drive produces a \(U_F\) that induces an intrinsically directed graph due to asymmetric transition probabilities. As a result, while graph connectivity provides a reliable diagnostic in the continuous drive case, it fails to capture the heating transition in the square-wave protocol, underscoring the crucial role of directionality in the underlying dynamics.

\section*{Acknowledgments}

We would like to thank Diptarka Das, Tapobrata Sarkar, Kunal Pal and Kuntal Pal for useful comments and valuable insights. Ankit Gill is thankful for the financial support received from the FARE Fellowship at IIT Kanpur.


\appendix

\begin{centering}
 \section*{Appendices}
\end{centering}

\section{$G_n$ analysis}  \label{app:Gn}

To analyze the autocorrelation function $G_n$, we first need to discuss the behaviour of $|\gamma_1|, \, |\gamma_2|$ and $\eta$ in different phases.

For brevity, define
\begin{equation}
	\psi_0 := \frac{\pi T_0}{L},
	\qquad
	\psi_1 := \frac{\pi T_1}{L},
\end{equation}
and then introduce:
\begin{equation}
	A := 2i(\sin\psi_0+\psi_1\cos\psi_0) = i a_0, \qquad B = 2(\cos\psi_0-\psi_1\sin\psi_0)\in\mathbb{R},
\end{equation}
where, $a_0 := 2(\sin\psi_0+\psi_1\cos\psi_0)\in\mathbb{R}$. Then in terms of these quantities, one can express,
\begin{equation}
	\Delta = (\sin\psi_0+\psi_1\cos\psi_0)^2-\psi_1^2 = \frac{a_0^2}{4}-\psi_1^2, \qquad \gamma_{1,2} = P(A \mp S), \qquad \eta = \frac{B+S}{B-S},
\end{equation}
with,
\begin{equation}
	P = -i\frac{L e^{-i\psi_0}}{2\pi T_1},
	\qquad
	S := \sqrt{-4 \Delta}.
\end{equation}
To analyze further, note the following identity:
\begin{align} \label{eq: gamma compare}
	|A+S|^2-|A-S|^2 = 4\,\Re(A\overline{S}) = 8 a_0 \, \Re( \sqrt{\Delta}) .
\end{align}
Then the following three cases can arise:

\subsubsection*{Case I: $\Delta < 0$ (heating phase)}

From \eqref{eq: gamma compare}, $\Re(A\overline{S}) = 0$. Hence,
\begin{equation}
	|A+S| = |A-S| ~~ \Rightarrow ~~ |\gamma_1| = |\gamma_2|,
\end{equation}
and $\eta$ is real,
\begin{align}
	\eta = \frac{B + 2  \sqrt{|\Delta|}}{B - 2  \sqrt{|\Delta|}} ~~ \Rightarrow ~~ \eta = e^{ \phi}, \, \phi \in \mathbb{R}.
\end{align}

\subsubsection*{Case II: $\Delta = 0$ (critical phase)}

In this phase, $S = 0$. Hence, 
\begin{align}
	|\gamma_1| = |\gamma_2|, \, \text{and,} \, \eta = 1.
\end{align}

\subsubsection*{Case III: $\Delta > 0$ (non-heating phase)}

For this case, $S = 2 i \sqrt{\Delta}$ and $\Re(A\overline{S}) = 8 a_0 \sqrt{\Delta}$. Hence following the expression of $a_0$, one can conclude:
\begin{equation} \label{eq:non-heating regimes}
	|\gamma_2| \gtrless |\gamma_1| \quad \text{for,} \quad \sin\frac{\pi T_0}{L} + \frac{\pi T_1}{L} \cos\frac{\pi T_0}{L} \gtrless 0.
\end{equation}
Also note, $\gamma_{1,2} = i P (a_0 \mp 2 \sqrt{\Delta})$, and thus, $\arg (\gamma_1) = \arg (\gamma_2)$. For this phase, $\eta$ lies on the unit circle:
\begin{align}
	\eta = \frac{B+2 i \sqrt{\Delta}}{B-2 i \sqrt{\Delta}} ~ \Rightarrow ~ \eta = e^{i \phi}, \, \phi = 4 \arctan \frac{\sqrt{\Delta}}{B} \in \mathbb{R}.
\end{align}

\subsection{The heating phase}

For brevity, we define,
\begin{equation}
	s=\sin\psi_0,\quad c=\cos\psi_0,
\end{equation}
so that,
\begin{align}
	B=2(c-\psi_1 s),\qquad  \Delta=(s+\psi_1 c)^2-\psi_1^2 < 0, \qquad S=\sqrt{-4\Delta} > 0, \qquad B^2 - S^2 = 4.
\end{align}
As discussed above, in this phase $\eta$ is real, $\eta=e^{\pm\varphi}$ with $\varphi\in\mathbb R_{>0}$. The sign of the exponent depends on the sign of $\sin\psi_0$, for which the following two subcases can appear:

\medskip

\noindent\textbf{Case 1:} $ \sin\psi_0>0$ (equivalently $(2m+1)L > T_0 > 2mL$, $m\in\mathbb Z_{\ge0}$).\\
Since in this case $s>0$, starting from $\Delta < 0$ one finds,
\begin{equation}
	\frac{c}{s} < \frac{\psi_1^2-1}{2\psi_1} < \psi_1 \qquad \Rightarrow \qquad B = 2(c - \psi_1 s) < 0,
\end{equation}
and,
\begin{equation}
	B^2 = 4 + S^2 \qquad \Rightarrow \qquad |B| > S.
\end{equation}
Following these conditions, one can then conclude,
\begin{equation}
	0 < \frac{B+S}{B-S} < 1 \quad\Rightarrow\quad \eta = e^{-\varphi},\, \varphi>0.
\end{equation}
Hence from \eqref{eq:Gn} one obtains, for positive conformal weight $h$ and large $n$,
\begin{equation}\label{eq:Gn heating-case1}
	G_n
	\simeq \left|\frac{\gamma_1-\gamma_2}{\gamma_2}\right|^{4h} e^{-2 n h \varphi}.
\end{equation}

\medskip

\noindent\textbf{Case 2:} $ \sin\psi_0<0$ (equivalently $(2m+2)L > T_0 > (2m+1)L$, $m\in\mathbb Z_{\ge0}$).\\
Following the similar arguments as above, in this subcase one finds $B>0$ and $B>S$, hence
\begin{equation}
	\frac{B+S}{B-S} > 1 \quad\Rightarrow\quad \eta = e^{\varphi'},\qquad \varphi'>0.
\end{equation}
Again for positive $h$ and large $n$,
\begin{equation}\label{eq:Gn heating-case2}
	G_n
	\simeq \left|\frac{\gamma_1-\gamma_2}{\gamma_1}\right|^{4h} e^{-2 n h \varphi'}.
\end{equation}

\medskip

Thus, in the heating phase $\Delta<0$, for both subcases $G_n$ decays exponentially in $n$ (and is suppressed by the conformal weight $h$):
\begin{equation}  \label{eq:Gn heating both}
	G_n \simeq 
	\begin{cases}
		\displaystyle \left| \frac{\gamma _1-\gamma _2}{\gamma _2} \right|^{4 h} e^{-2 n h \varphi} & \text{for } (2m+1)L > T_0 > 2mL,\\[10pt]
		\displaystyle \left| \frac{\gamma _1-\gamma _2}{\gamma _1} \right|^{4 h} e^{-2 n h \varphi'} & \text{for } (2m+2)L > T_0 > (2m+1)L.
	\end{cases}
\end{equation}
In terms of $\eta$,
\begin{equation} 
	G_n \simeq 
	\begin{cases}
		\displaystyle \left| \frac{\gamma _1-\gamma _2}{\gamma _2} \right|^{4 h} |\eta|^{2 n h} & \text{for } (2m+1)L > T_0 > 2mL,\\[10pt]
		\displaystyle \left| \frac{\gamma _1-\gamma _2}{\gamma _1} \right|^{4 h} |\eta|^{-2 n h} & \text{for } (2m+2)L > T_0 > (2m+1)L.
	\end{cases}
\end{equation}

\subsection{The non-heating phase} \label{app:Gn nonheating}

As discussed earlier, in this phase $\eta = e^{i  \phi}$ and $\arg (\gamma_1) = \arg (\gamma_2)$. Hence the autocorrelation takes the form: 
\begin{equation}
	G_n \;=\; 
	\frac{|\gamma_1-\gamma_2|^{4h}}{\left((\gamma_1-\gamma_2 e^{i n\phi})(\overline{\gamma_1}-\overline{\gamma_2} e^{-i n\phi})\right)^{2h}} = \frac{|\gamma_1-\gamma_2|^{4h}}{\mathcal{A}-\mathcal{B}\cos(n\phi)} = \frac{|\gamma_1-\gamma_2|^{4h}}{\mathcal{D}(n\phi)} ,
\end{equation}
where, $\mathcal{A}:=|\gamma_1|^2+|\gamma_2|^2, \, 
\mathcal{B}:=2|\gamma_1||\gamma_2|, \, \mathcal{D}(n\phi) = \mathcal{A} - \mathcal{B} \cos(n\phi)$. The denominator in the above expression can be expanded further,
\begin{equation}
	\frac{1}{\mathcal{D}(n\phi)^{2h}} = \frac{(1-q^2)^{2h}}{\mathcal{C}^{2h}} \sum_{m=-\infty}^{\infty} q^{|m|}
	\sum_{\ell=0}^{\infty} \binom{2h+|m|+\ell-1}{|m|+\ell} \binom{2h+\ell-1}{\ell} q^{2\ell} \, e^{i m n\phi},
\end{equation}
where,
\begin{equation}
	\mathcal{C}:=\sqrt{\mathcal{A}^2-\mathcal{B}^2},
	\qquad 
	q:=\frac{\mathcal{B}}{\mathcal{A}+\mathcal{C}}
	\qquad (0<q<1).
\end{equation}
Now using the identity,
\begin{equation}
	\sum_{\ell=0}^{\infty} \binom{2h+m+\ell-1}{m+\ell} \binom{2h+\ell-1}{\ell} z^{\ell}
	= \binom{2h+m-1}{m}\; {}_2F_1\big(2h,2h+m;m+1;z\big),
\end{equation}
one can finally obtain:
\begin{equation}
	\begin{aligned}
		G_n
		&=
		|\gamma_1-\gamma_2|^{4h}
		\frac{(1-q^2)^{2h}}{\mathcal{C}^{2h}}
		\Bigg\{
		{}_2F_1(2h,2h;1;q^2)
		\\
		&\qquad
		+
		2\sum_{m=1}^{\infty}
		q^{m}
		\binom{2h+m-1}{m}
		{}_2F_1\!\big(2h,2h+m;m+1;q^2\big)
		\cos(m n\phi)
		\Bigg\}.
	\end{aligned}
\end{equation}
This can be further simplified depending on the two regimes mentioned in \eqref{eq:non-heating regimes}:
\begin{equation} 
	\begin{aligned}
		G_n
		&=
		(1-2r+r^2)^{2h}
		\Bigg\{
		{}_2F_1(2h,2h;1;r^2)
		\\
		&\qquad
		+
		2\sum_{m=1}^{\infty}
		r^{m}
		\binom{2h+m-1}{m}
		{}_2F_1\!\big(2h,2h+m;m+1;r^2\big)
		\cos(m n\phi)
		\Bigg\},
	\end{aligned}
\end{equation}
where,
\begin{equation}
	r=
	\begin{cases}
		\dfrac{|\gamma_2|}{|\gamma_1|}  ~~~ \text{for } |\gamma_1| > |\gamma_2|, \, \text{i.e., } \sin\frac{\pi T_0}{L} + \frac{\pi T_1}{L} \cos\frac{\pi T_0}{L} < 0, \\[10pt]
		\dfrac{|\gamma_1|}{|\gamma_2|}  ~~~ \text{for } |\gamma_2| > |\gamma_1|, \, \text{i.e., } \sin\frac{\pi T_0}{L} + \frac{\pi T_1}{L} \cos\frac{\pi T_0}{L} > 0.
	\end{cases}
\end{equation}

\bibliography{refs} 
\bibliographystyle{jhep}

\end{document}